\documentclass[12pt,a4paper]{article}

\usepackage{amsmath,amssymb}
\usepackage[dvips]{lscape,graphicx}

\newcommand{\sfrac}[2]{\frac{\scriptstyle #1}{\scriptstyle #2}}

\begin{document}

\author{R. B. Nevzorov and M. A. Trusov \\ {\itshape ITEP, Moscow, Russia}}

\title{Infrared quasi--fixed solutions in the NMSSM}

\maketitle

\begin{abstract}
The considerable part of the parameter space in the MSSM
corresponding to the infrared quasi fixed point scenario is almost
excluded by LEP\,II bounds on the lightest Higgs boson mass. In
the NMSSM the mass of the lightest Higgs boson reaches its maximum
value in the strong Yukawa coupling limit when Yukawa couplings
are essentially larger than gauge ones at the Grand Unification
scale. In this limit the solutions of the renormalisation group
equations are attracted to the infrared and Hill type effective
fixed lines or surfaces in the Yukawa coupling parameter space.
They are concentrated in the vicinity of quasi fixed points for
$Y_i(0)\to\infty$. However the solutions are attracted to such
points rather weakly. For this reason when all $Y_i(0)\sim 1$ the
solutions of the renormalisation group equations are gathered near
a line in the Hill type effective surface. In the paper the
approximate solutions for the NMSSM Yukawa couplings are given.
The possibility of $b$--quark and $\tau$--lepton Yukawa coupling
unification at the scale $M_{X}$ is also discussed.
\end{abstract}

\newpage

\section{Introduction}

The existence of quasi--fixed points is among the most spectacular
and the most interesting properties of renormalization group
equations. A feature characteristic to those solutions to
renormalization group equations that approach such points is that
a number of fundamental parameters of the theory are focused in a
narrow interval in the infrared region. This means that, at the
electroweak scale, some constants or their combinations cease to
depend on the boundary conditions. That solutions to
renormalization group equations behave in so peculiar a way in the
vicinities of quasi--fixed points results in that the parameter
space of the theory being considered is constrained for a wide
class of such solutions. As a result, the predictive power of the
theories being discussed becomes higher near these points.
Nonetheless, it turns out that, within the minimal Standard Model
(SM), the quasi--fixed point scenario leads to overly high a value
for the mass of the $t$--quark, which contradicts experimental
data obtained at FNAL.

In contrast to the SM, its supersymmetric (SUSY) generalization --
the Minimal SUSY Standard Model (MSSM) -- features two Higgs
doublets (not one), $H_1$ and $H_2$. Upon a spontaneous breakdown
of symmetry, they develop nonzero vacuum expectation values $v_1$
and $v_2$, with the constraint
$v^2=v_1^2+v_2^2=(246\text{~GeV})^2$ being satisfied. In relation
to what occurs in the SM, the $t$--quark running mass $m_t$ that
is generated within SUSY models upon the breakdown of
$SU(2)\otimes U(1)$ gauge symmetry involves an additional factor
$\sin\beta$,
\begin{equation}
m_t(M_t^{\text{pole}})=\frac{h_t(M_t^{\text{pole}})}{\sqrt{2}}v\sin\beta,
\label{1}
\end{equation}
where $\tan\beta=v_2/v_1$ and $h_t$ is the Yukawa coupling
constant for the $t$--quark. Since $\sin\beta\le 1$,
$m_t(M_t^{\text{pole}})$ is always less in the MSSM than in the SM
at the same values of the Yukawa coupling constants. Recent
experimental data on the $t$--quark mass make it possible to
determine $m_t(M_t^{\text{pole}})$ within the $\overline{MS}$
scheme \cite{1}. It proves to be $m_t(M_t^{\text{pole}})=165\pm
5\text{~GeV}$. The uncertainty in the determination of the running
mass of the $t$--quark stems predominantly from the experimental
error with which its pole mass was measured
($M_t^{\text{pole}}=174.3\pm 5.1\text{~GeV}$ \cite{2}).

Equation (\ref{1}) unambiguously relates $\tan\beta$ to the value
of the Yukawa coupling constant for the $t$--quark at the
electroweak scale. At modest values of $\tan\beta$ ($\tan\beta\ll
50-60$), the Yukawa coupling constants for the $b$--quark, $h_b$,
and for the $\tau$--lepton, $h_\tau$, are negligibly small, which
makes it possible to obtain an analytic solution to the
renormalization group equation within the MSSM \cite{3}. In this
case, the boundary conditions are imposed at the scale $M_X\approx
3\cdot 10^{16}\text{~GeV}$, where the gauge coupling constants are
naturally unified within the MSSM. For the $t$--quark Yukawa
coupling constant, it is convenient to represent an exact solution
to the renormalization group equations in the form
\begin{equation}
Y_t(t)=\frac{\dfrac{E(t)}{6F(t)}}{1+\dfrac{1}{6Y_t(0)F(t)}}\, ,
\label{2}
\end{equation}
where $Y_t(t)=h_t^2(t)/(4\pi)^2$ and $t=\ln(M_X^2/q^2)$. The
explicit expressions for the functions $E(t)$ and $F(t)$ are
presented in the Appendix (see (\ref{A.4})). At the electroweak
scale, the second term in parentheses is much less than unity for
$h_t^2(0)\ge 1$. The dependence of $h_t^2(t)$ on the initial
conditions at $t=0$ is weak, and the relevant solution to the
renormalization group equations approaches a quasi--fixed point
\cite{4}: $Y_{\text{QFP}}(t)=E(t)/6F(t)$. Formally, a solution of
this type can be obtained by making $Y_t(0)$ tend to infinity in
expression (\ref{2}). The situation here is, however, different
from that near the Pendleton--Ross infrared fixed point
\cite{5}-\cite{8}, which solutions to the renormalization group
equations approach only in the asymptotic regime for $q^2\to 0$:
the deviation from $Y_{\text{QFP}}$ at finite values of $Y_t(0)$
is determined by the ratio $Y_{\text{QFP}}(t)/(E(t)Y_t(0))$, which
is of order $1/(10h_t^2(0))$ at the electroweak scale and which is
small at comparatively large $h_t^2(0)$ ($h_t^2(0)\ge 1$). For a
wide class of solutions, this interesting property of the
renormalization group equations within the MSSM makes it possible
to predict quite precisely the value of the Yukawa coupling
constant for the $t$--quark at the scale $q=M_t^{\text{pole}}$,
\begin{equation}
h_{\text{QFP}}^2(t_0)=0.87\cdot g_3^2(t_0)=1.26, \label{3}
\end{equation}
where $g_3$ is the gauge coupling constant for strong interaction
and $t_0=2\ln(M_X/M_t^{\text{pole}})$. The accuracy of this
prediction becomes higher with increasing $h_t^2(0)$. At
sufficiently large initial values of $Y_t(t)$, it would be
illegitimate to restrict the analysis to one--loop renormalization
group equations -- it is necessary to take into account higher
order perturbative corrections. Moreover, the value of the Yukawa
coupling constant for the $t$--quark at the electroweak scale
depends on the strong interaction coupling constant, which we set
to $\alpha_3(M_Z)=0.118$. Nevertheless, all these uncertainties do
not lead to significant deviation from (\ref{3}). By way of
example, we indicate that the calculations that were performed in
\cite{9} and which employed the four--loop beta function showed
that deviations from (\ref{3}) are within $2\%$.

For each fixed value of $Y_t(0)$, the Yukawa coupling constant for
the $t$--quark at the electroweak scale can be evaluated by using
the exact analytic solution (\ref{2}), whereupon $\tan\beta$ can
be determined by substituting the resulting value of $h_t(t_0)$
into (\ref{4}). The theoretical analysis performed in
\cite{10}-\cite{13} revealed that, for the renormalization group
equations within the MSSM, a broad class of solutions
corresponding to the infrared quasi--fixed point regime leads to
$\tan\beta$ values ranging between $1.3$ and $1.8$. With
increasing Yukawa coupling constant for the $t$--quark, the
corresponding trilinear coupling constant $A_t$ for the
interaction of scalar particles and the combination
$\mathfrak{M}_t^2=m_Q^2+m_U^2+m_2^2$ of the scalar particle masses
cease to depend on the initial conditions. In the vicinity of the
quasi--fixed point, they are expressed in terms of only the
gaugino mass at the scale $M_X$, with the result that the
parameter space is further constrained. In the infrared
quasi--fixed point regime at $\tan\beta\sim 1$, the properties of
solutions to the set of renormalization group equations and the
spectrum of particles were investigated in \cite{8},
\cite{12}-\cite{16}.

Finally, there is yet another circumstance that appears as an
incentive to study the limit of strong Yukawa coupling within the
MSSM. Minimal schemes that are used to unify gauge interactions
and which are based on gauge groups like $SU(5)$, $E_6$, or
$SO(10)$ predict the equality of the Yukawa coupling constants
$h_b$ and $h_\tau$ for , respectively, the $b$--quark and the
$\tau$--lepton at the scale $M_X$ \cite{17}. Within the MSSM,
$h_b$ and $h_\tau$ are unified at two specific values of
$h_t(M_t^{\text{pole}})$. One of these fails within a narrow
region near $h_{\text{QFP}}(t_0)$, while the other corresponds to
the scenario of large $\tan\beta$. In more detail, the problem of
$b-\tau$ unification within the MSSM was discussed in \cite{7},
\cite{15},\cite{16},\cite{18},\cite{19},\cite{20}.

The spectrum of the Higgs sector of the MSSM contains four massive
states: two CP--odd states, one CP--even state, and one charged
state. The presence of a light Higgs boson in the CP--even sector
is an important feature of SUSY models. The upper limit on its
mass greatly depends on $\tan\beta$. A reduction of the number of
independent parameters in the infrared quasi--fixed point regime
made it possible to determine, to a sufficiently high degree of
precision, an upper limit on the mass of the lightest CP--even
Higgs boson. In the case being considered, comparatively small
values of $\tan\beta$ result in that its mass does not exceed
$94\pm 5\text{~GeV}$ \cite{11}-\cite{13}. This limit is
$25-30\text{~GeV}$ lower than the absolute upper limit in the
minimal SUSY model. At the same time the lower limit on the mass
of the lightest Higgs boson from LEP\,II data -- in the case of a
heavy spectrum of SUSY particles, it coincides with the
corresponding limit on the Higgs boson mass in the SM -- is
$113.3\text{~GeV}$ \cite{21}. Actually, this means that a major
part of solutions approaching the infrared quasi--fixed point
within the MSSM have already been ruled out by the existing
LEP\,II data. In order to meet the experimental constraints on the
mass of the lightest Higgs boson, it is necessary either to go
over to studying solutions that lead to large values of
$\tan\beta$ within the MSSM or to extend the Higgs sector of the
minimal SUSY model. The detailed investigations that were
performed in \cite{13}, \cite{16}, \cite{19}, \cite{22} revealed
that, at $\tan\beta\approx 50-60$, solutions to the
renormalization group equations also approach the infrared
quasi--fixed point, the basic properties of the solutions
remaining unchanged.

The nonminimal SUSY SM (NMSSM) \cite{23}-\cite{25} whose Higgs
sector contains, in addition to the doublets $H_1$ and $H_2$, an
extra superfield $Y$ that is a singlet with respect to
$SU(2)\otimes U(1)$ gauge interactions is the simplest extension
of the MSSM. In the parameter space of the NMSSM, the region that
corresponds to the limit of strong Yukawa coupling, in which case
the Yukawa coupling constants $Y_i(0)$ at the Grand Unification
scale $M_X$ are much greater than the gauge coupling constant
$\tilde{\alpha}(0)$, is that which is the most appealing from the
point of view of a theoretical analysis. It is the region where
the upper limit on the mass of the lightest Higgs boson takes a
maximum value that is few GeV greater than the corresponding
absolute limit within the MSSM \cite{26}. Moreover, it is
possible, in the case being considered, to choose coupling
constants in such a way as to obtain the unification of the Yukawa
coupling constants for the $b$--quark and the $\tau$--lepton at
the scale $M_X$.

For the Yukawa coupling constants in the limit of strong Yukawa
coupling, we study here basic properties of solutions to the
renormalization group equations within the NMSSM. We show that, in
the limit $Y_i(0)\to\infty$, all solutions in the nonminimal SUSY
model are concentrated, as in the MSSM, near quasi--fixed points
that arise as the result of intersections of Hill lines or
surfaces with some invariant line in the space of Yukawa
coordinates. However, the solution are rather weakly attracted to
these points. For $Y_i(0)\gg\tilde{\alpha}(0)$, all solutions to
the renormalization group equations are therefore nonuniformly
distributed near Hill lines or surfaces. Approximate solutions to
the set of nonlinear differential equations that describe the
evolution of $Y_i(t)$ within the NMSSM are presented in the
Appendix. The approximate solutions that are obtained in the
present study are compared with the results of numerical
calculations within the nonminimal SUSY model.

\section{Upper limit on the mass of the lightest Higgs boson and renormalisation group equations in the NMSSM}

By construction, the superpotential of the NMSSM is invariant
under the discrete transformations $y'_\alpha=e^{2\pi
i/3}y_\alpha$ of the $Z_3$ group \cite{24}. The term $\mu(H_1H_2)$
in the superpotential of the NMSSM does not satisfy this
requirement. For this reason, an extra superfield $Y$ that is a
singlet with respect to $SU(2)\otimes U(1)$ gauge interactions is
introduced in the NMSSM. The superpotential of the Higgs sector of
the NMSSM \cite{23}-\cite{25} has the form
\begin{equation}
W_{h}=\lambda Y(H_1 H_2)+\frac{\varkappa}{3}Y^3. \label{4}
\end{equation}
Upon a spontaneous breakdown of $SU(2)\otimes U(1)$ symmetry, the
field $Y$ develops a nonzero vacuum expectation value ($\langle
Y\rangle=y/\sqrt{2}$) and there arises an effective $\mu$--term
($\mu=\lambda y/\sqrt{2}$).

The introduction of the neutral field $Y$ in the superpotential of
the NMSSM leads to the emergence of the corresponding $F$--term in
the potential of the interaction of Higgs fields. As a result, the
upper limit on the mass of the lightest Higgs boson proves to be
greater than in the MSSM. Specifically, we have
\begin{equation}
m_h\le\sqrt{\frac{\lambda^2}{2} v^2\sin^2 2\beta+ M_Z^2\cos^2
2\beta+\Delta_1+\Delta_2}, \label{5}
\end{equation}
where $\Delta_1$ and $\Delta_2$ stand for, respectively, one--loop
and two--loop corrections. At $\lambda=0$, the expressions for the
above upper limit within the MSSM and the NMSSM coincide. In the
tree approximation, relation (\ref{5}) was obtained in \cite{25}.
The inclusion of loop corrections to the effective potential of
Higgs fields leads to a considerable growth of the upper limit on
$m_h$. The main contributions to $\Delta_1$ and $\Delta_2$ come
from loops involving a $t$--quark and its superpartners. In the
leading approximation, the contribution of loop corrections to the
upper limit on the Higgs boson mass within the NMSSM is
approximately equal to that within the minimal SUSY model. In
calculating the corrections $\Delta_1$ and $\Delta_2$ within the
NMSSM, it is necessary, however, to replace the parameter $\mu$ by
$\lambda y/\sqrt{2}$. One--loop and two--loop corrections within
the MSSM were studied in \cite{27} and \cite{28}, respectively. In
the leading approximation, these corrections are proportional to
$m_t^4$; they depend logarithmically on the scale of SUSY
breaking, $M_S=\sqrt{m_{\tilde{t}_1}m_{\tilde{t}_2}}$
($m_{\tilde{t}_1}$ and $m_{\tilde{t}_2}$ are the masses of the
superpartners of the $t$--quark), and are virtually independent on
the choice of $\tan\beta$. The Higgs sector in the nonminimal SUSY
model and one--loop corrections to this sector were studied in
\cite{29}-\cite{31}. The possibility of a spontaneous
CP--violation in the Higgs sector of the NMSSM was considered in
\cite{31},\cite{32}. In \cite{33}, the upper limit on the mass of
the lightest Higgs boson within the NMSSM was compared with the
corresponding limits within the minimal SM and minimal SUSY
models. The most recent investigations revealed that, in the
nonminimal SUSY model, $m_h$ does not exceed 135~GeV \cite {26}.

From relation (\ref{5}), it follows that the upper limit on $m_h$
grows with the increasing $\lambda(t_0)$. It should be emphasised
that only in the region of small $\tan\beta$ is this limit
markedly different from the corresponding limit within the MSSM.
At large values of this parameter ($\tan\beta\gg 1$), the quantity
$\sin 2\beta$ vanishes, so that the upper limits on the mass of
the lightest Higgs boson within the MSSM and the NMSSM virtually
coincide. But only in the case of sufficiently large $h_t(t_0)$ is
the scenario of small $\tan\beta$ realised, $\tan\beta$ becoming
smaller with increasing $h_t(t_0)$, as can be seen from relation
(\ref{1}). At the same time, an analysis of the renormalisation
group equations within the MSSM and the NMSSM reveals that the
growth of the Yukawa coupling constants at the electroweak scale
is accompanied by an increase in $h_t(0)$ and $\lambda(0)$ at the
Grand Unification scale. Thus, it becomes clear that the upper
limit on the mass of the lightest Higgs boson within the
nonminimal SUSY model attains a maximum value in the limit of
strong Yukawa coupling, in which case
$Y_t(0),Y_\lambda(0)\gg\tilde{\alpha}(0)$.

From the point of view of a renormalisation group analysis,
investigation of the NMSSM presents a much more complicated
problem than investigation of the minimal SUSY model. The full set
of renormalization group equations within the NMSSM can be found
in \cite{30}, \cite{34}. Even in the one--loop approximation, this
set of equations os nonlinear and its analytic solution does not
exist. All equations forming this set can be partitioned into two
groups, the first containing equations that describe the evolution
of gauge and Yukawa coupling constants. In analysing the nonlinear
differential equations entering into this group, it is convenient
to go over from $h_t$, $\lambda$, and $\varkappa$ to the
quantities $\rho_t$, $\rho_\lambda$, and $\rho_\varkappa$, which
are defined as the ratios of the squares of the corresponding
Yukawa coupling constants and the gauge coupling constant for
strong interaction,
\[ \rho_t(t)=\frac{Y_t(t)}{\tilde{\alpha}_3(t)},\quad
\rho_{\lambda}(t)=\frac{Y_{\lambda}(t)}{\tilde{\alpha}_3(t)},\quad
\rho_{\varkappa}(t)=\frac{Y_{\varkappa}(t)}{\tilde{\alpha}_3(t)},
\]
where $\tilde{\alpha}_3(t)=g^2_3(t)/(4\pi)^2$,
$Y_t(t)=h^2_t(t)/(4\pi)^2$,
$Y_{\lambda}(t)=\lambda^2(t)/(4\pi)^2$, and
$Y_{\varkappa}(t)=\varkappa^2(t)/(4\pi)^2$. The one--loop
renormalisation group equations for $\rho_i(t)$ have the form
\begin{eqnarray}
\frac{d\tilde{\alpha}_3}{dt}&=&3\tilde{\alpha}_3^2 \nonumber \\
\frac{d\rho_1}{dt}&=&-\tilde{\alpha}_3\rho_1\left(
\frac{33}{5}\rho_1+3\right) \nonumber \\
\frac{d\rho_2}{dt}&=&-\tilde{\alpha}_3\rho_2\left( \rho_2+3\right)
\nonumber \\
\frac{d\rho_t}{dt}&=&-\tilde{\alpha}_3 \rho_t\left(
6\rho_t+\rho_{\lambda}-\frac{7}{3}-3\rho_2-\frac{13}{15}
\rho_1\right) \label{6} \\
\frac{d\rho_{\lambda}}{dt}&=&-\tilde{\alpha}_3\rho_{\lambda}
\left(3\rho_t+4\rho_{\lambda}+2\rho_{\varkappa}+3-3\rho_2-
\frac{3}{5}\rho_1\right) \nonumber \\
\frac{d\rho_{\varkappa}}{dt}&=&-\tilde{\alpha}_3 \rho_{\varkappa}
\left(6\rho_{\lambda}+6\rho_{\varkappa}+3\right), \nonumber
\end{eqnarray}
where $\rho_1(t)=\tilde{\alpha}_1(t)/\tilde{\alpha}_3(t)$,
$\rho_2(t)=\tilde{\alpha}_2(t)/\tilde{\alpha}_3(t)$,
$\tilde{\alpha}_1(t)=g_1^2(t)/(4\pi)^2$, and
$\tilde{\alpha}_2(t)=g_2^2(t)/(4\pi)^2$. The second group includes
equations for the parameters of a soft breakdown of SUSY, which
are necessary for obtaining a phenomenologically acceptable
spectrum of superpartners of observable particles. Since boundary
conditions for three Yukawa coupling constants are unknown, it is
very difficult to perform a numerical analysis of the equations
belonging to the first group and of the full set of the equations
given above. In the regime of strong Yukawa coupling, however,
solutions to the renormalisation group equations are concentrated
in a narrow region of the parameter space near the electroweak
scale, and this considerably simplifies the analysis of the set of
equations being considered.

\section{Invariant and quasi--fixed lines; a determination of the quasi--fixed point}

Let us first consider the simplest case of $\varkappa=0$. The
growth of the Yukawa coupling constant $\lambda(t_0)$ at a fixed
value of $h_t(t_0)$ results in that the Landau pole in solutions
to the renormalization group equations approaches the Grand
Unification scale from above. At a specific value
$\lambda(t_0)=\lambda_{\text{max}}$, perturbation theory at $q\sim
M_X$ cease to be applicable. With increasing (decreasing) Yukawa
coupling constant for the $b$--quark, $\lambda_{\text{max}}$
decreases (increases). In the $(\rho_t,\rho_\lambda)$ plane, the
dependence $\lambda^2_{\text{max}}(h_t^2)$ is represented by a
curve bounding the region of admissible values of the parameters
$\rho_t(t_0)$ and $\rho_\lambda(t_0)$. At $\rho_\lambda=0$, this
curve intersects the abscissa at the point
$\rho_t=\rho_t^{\text{QFP}}(t_0)$. This is the way in which there
arises, in the $(\rho_t,\rho_\lambda)$ plane, the quasi--fixed (or
Hill) line near which solutions to the renormalization group
equations for the initial values of the Yukawa coupling constants
in the range $2\le h_t^2(0),\lambda^2(0)\le 10$ are grouped (see
Figs. 1a, 1b). With increasing $\lambda^2(0)$ and $h_t^2(0)$, the
region where the solutions in questions are concentrated sharply
shrinks. At initial values of the Yukawa coupling constants from
the range between $20$ and $100$, they are grouped in a narrow
region near the straight line
\begin{equation}
\rho_t(t_0)+0.506\rho_{\lambda}(t_0)=0.91, \label{7}
\end{equation}
which can be obtained by fitting the results of numerical
calculations (these results are presented in Figs. 2a and 2b).
Moreover, it follows from the data in the Table that the
combination $h_t^2(t_0)+0.506\lambda^2(t_0)$ of the Yukawa
coupling constants depends much more weakly on $\lambda^2(0)$ and
$h_t^2(0)$ than $\lambda^2(t_0)$ and $h_t^2(t_0)$ individually. In
other words, a decrease in $\lambda^2(t_0)$ compensates for an
increase in $h_t^2(t_0)$, and vice versa. To illustrate this, we
indicate that, at initial values $\lambda^2(0)$ and $h_t^2(0)$
from the interval $(2,10)$, the following occurs upon an increase
in $\lambda^2(0)$ and a decrease in $h_t^2(0)$: the constant
$\lambda^2(t_0)$ increases monotonically from $0.191$ to $0.421$,
while $h_t^2(t_0)$ decreases from $1.199$ to $1.051$; at the same
time, the sum $h_t^2(t_0)+0.506\lambda^2(t_0)$ at identical
$\lambda^2(0)$ and $h_t^2(0)$ ranges between $1.266$ and $1.318$.
The results in Figs. 3a and 3b, which illustrate the evolution of
the above combinations of the Yukawa coupling constants, also
confirm that this combination is virtually independent of the
initial conditions.

In analysing the results of numerical calculations, our attention
is engaged by a pronounced nonuniformity in the distribution of
solutions to the renormalization group equations along the
infrared quasi--fixed line. The main reason for this is that, in
the regime of strong Yukawa coupling, the solutions in question
are attracted not only to the quasi--fixed but also to the
infrared fixed (or invariant) line. The latter connects two fixed
points. Of these, one is an infrared fixed point of the set of
renormalization group equations within the NMSSM ($\rho_t=7/18$,
$\rho_{\lambda}=0$, $\rho_1=(\tilde{\alpha}_1/
\tilde{\alpha}_3)=0$,
$\rho_2=(\tilde{\alpha}_2/\tilde{\alpha}_3)=0$) \cite{6}, while
the other fixed point $(\rho_\lambda/\rho_t=1)$ corresponds to
values of the Yukawa coupling constants in the region
$Y_t,Y_\lambda\gg\tilde{\alpha}_i$, in which case the gauge
coupling constants on the right--hand sides of the renormalization
group equations can be disregarded \cite{35}. The infrared fixed
line is invariant under renormalization group transformations --
that is, it is independent of the scale at which the boundary
values $Y_t(0)$ and $Y_\lambda(0)$ are specified and of the
boundary values themselves. If the boundary conditions are such
that $Y_t(0)$ and $Y_\lambda(0)$ belong to the fixed line, the
evolution of the Yukawa coupling constants proceeds further along
this line toward the infrared fixed point of the set of
renormalization group equations within the NMSSM. With increasing
$t$, all other solutions to the renormalization group equations
are attracted to the infrared fixed line and, for $t/(4\pi)\gg 1$,
approach the stable infrared fixed point. From the data in Figs.
1b and 2b, it follows that, with increasing $Y_t(0)$ and
$Y_\lambda(0)$, all solutions to the renormalization group
equations are concentrated in the vicinity of the point of
intersection of the infrared fixed and the quasi--fixed line:
\[
\rho^{\text{QFP}}_t(t_0)=0.803,\qquad \rho^{\text{QFP}}_{\lambda}(t_0)=0.224.
\]
Hence, this point can be considered as the quasi--fixed point of
the set of renormalization group equations within the NMSSM at
$\varkappa=0$.

Infrared fixed lines and surfaces, as well as their properties in
the minimal Standard Model and in the minimal SUSY model, were
studied in detail in \cite{7}, \cite{20}, \cite{36}. Within the
NMSSM, the emergence of fixed lines can be traced at $\lambda=0$,
in which case the set of renormalization group equations for the
Yukawa coupling constants reduced to two independent differential
equations -- of these, one coincides with the equation for
$Y_t(t)$ in the minimal SUSY model, while the other describes the
evolution of $Y_\varkappa(t)$. In the limit being considered, the
set of one--loop renormalization group equations has the exact
analytic solution
\begin{eqnarray}
Y_{\varkappa}(t)&=&\frac{Y_{\varkappa}(0)}{1+6Y_{\varkappa}(0)t},
\nonumber \\
Y_t(t)&=&\frac{Y_t(0)E(t)}{1+6Y_t(0)F(t)}, \label{8}
\\
\tilde{\alpha}_i(t)&=&\frac{\tilde{\alpha}_i(0)}{1+b_i\alpha_i(0)t},
\nonumber
\end{eqnarray}
where the expressions for $E(t)$, $F(t)$, and $b_i$ are presented
in the Appendix. The quasi--fixed line in the
$(\rho_t,\rho_\varkappa)$ plane includes two straight lines
parallel to he coordinate axes (see Fig. 4b),
\begin{equation}
\begin{split}
\rho_t&=\frac{E(t_0)}{6\tilde{\alpha}_3(t_0)F(t_0)}\approx
0.876,\\
\rho_{\varkappa}&=\frac{1}{6\tilde{\alpha}_3(t_0)t_0}\approx
0.280,
\end{split}
\label{9}
\end{equation}
which intersects at the point $(0.876,0.280)$. Since the above
solutions to the renormalization group equations are attracted to
the invariant line at $t/(4\pi)\gg 1$, unity can be disregarded in
the denominators of $Y_t(t)$ and $Y_\varkappa(t)$. The infrared
fixed line can then be specified parametrically:
\begin{equation}
\begin{split}
\rho_t(t)&=\frac{E(t)}{6\tilde{\alpha}_3(t)F(t)} \\
\rho_{\varkappa}(t)&=\frac{1}{6\tilde{\alpha}_3(t)t}\, .
\end{split}
\label{10}
\end{equation}
It can easily be shown that the limit $t\to 0$ corresponds to the
values $\rho_t$ and $\rho_\varkappa\gg 1$ belonging to this curve,
$\rho_t$ and $\rho_\varkappa$ being virtually coincident in this
limit. By using the expansions of the functions $E(t)$ and $F(t)$
in the vicinity of the origin, $F(t)\approx t+0.5E'(0)t^2+\dots$
and $E(t)\approx 1+E'(0)t+\dots$, we obtain
\begin{equation}
\rho_{\varkappa}=\rho_t-\frac{4}{9}-\frac{\rho_2}{4}-
\frac{13}{180}\rho_1. \label{11}
\end{equation}
The equality $\rho_\varkappa=\rho_t$ corresponds to the stable
fixed point of the renormalization group equations in the regime
of strong Yukawa coupling ($\rho_\varkappa,\rho_t\gg 1$). As
$\tilde{\alpha}_3$ tends to the Landau pole for $t\to
t_c=1/(3\tilde{\alpha}_3(0))$, however, the line given by
Eq.(\ref{10}) approaches the stable infrared point (see Fig. 4b);
that is, $\rho_t(t)$ tends to $7/18$, while $\rho_\varkappa(t)$
vanishes: $\rho_\varkappa\sim(\rho_t-7/18)^{9/7}$. The curve given
by (\ref{10}), which connects the fixed points
$\rho_\varkappa/\rho_t=1$ and $\rho_\varkappa=0,~\rho_t=7/18$
intersects the quasi--fixed line at the point $(0.876,0.280)$. As
can be seen from Fig. 4b, solutions to the renormalization group
equations are concentrated precisely in the vicinity of this
point.

Near the infrared fixed point, the curve being investigated is
tangent to another invariant line, that which is specified by the
equation $\rho_\varkappa=0$. This line connects the unstable fixed
point $\rho_\varkappa/\rho_t=0$, which arises in the regime of
strong Yukawa coupling ($\rho_t\gg 1$), with other fixed points,
those at $\rho_\varkappa=0,~\rho_t=7/18$ and at
$\rho_\varkappa=\rho_t=0$, the last also being unstable. Yet
another infrared fixed line -- the attraction of solutions to the
renormalization group equations to this line os the weakest --
passes through the points $\rho_t/\rho_\varkappa=0$ and
$\rho_t=7/18~\rho_\varkappa=0$. At
$\tilde{\alpha}_1=\tilde{\alpha}_2=0$, it appears to be a straight
line parallel to the coordinate axis, $\rho_t=7/18$. However, the
inclusion of electroweak interactions leads to a monotonic
decrease in $\rho_t(t)$ with increasing $\rho_\varkappa(t)$. In
the vicinity of the stable infrared fixed point for $t\to t_c$,
the equation for this line has the form
\begin{equation}
\begin{split}
\rho_t(t)&=\frac{7}{18}-\frac{7}{4}\rho_2(t)-\frac{91}{180}\rho_1(t)
\\ \rho_{\varkappa}(t)&=\frac{1}{6\tilde{\alpha}_3(t)t}\, .
\end{split}
\label{12}
\end{equation}

Apart from the replacement of $\rho_\varkappa$ by $\rho_\lambda$,
the same infrared fixed lines are involved in the analysis of
renormalization group equations within the NMSSM in the case where
$\varkappa=0$. As before, the invariant line that connects the
stable fixed points $\rho_\lambda/\rho_t=1$ and
$\rho_t=7/18,~\rho_\lambda=0$ attracts most strongly solutions to
the renormalization group equations. Nevertheless, the asymptotic
behaviour of the curve being studied changes for
$\rho_\lambda,\rho_t\gg 1$, where it becomes
\begin{equation}
\rho_{\lambda}=\rho_t-\frac{8}{15}-\frac{2}{75}\rho_1, \label{13}
\end{equation}
and in the vicinity of the point $\rho_t=7/18, ~\rho_\lambda=0$,
where we have $\rho_\lambda\sim(\rho_t-7/18)^{25/14}$. In
analysing the behaviour of solutions to the renormalization group
equations, the other two invariant lines have but a marginal
effect. One of these is specified by the equations
$\rho_\lambda=0$. The second connects the unstable fixed point in
the regime of strong Yukawa coupling, $\rho_t/\rho_\lambda=0$,
with the stable infrared point, near which we have
$\rho_\lambda\sim(7/18-\rho_t)^{25/18}$.

\section{Invariant and Hill surfaces}

In a more complicated case where all three Yukawa coupling
constants in the NMSSM are nonzero, analysis of the set of
renormalization group equations presents a much more difficult
problem. In particular, invariant (infrared fixed) and Hill
surfaces come to the fore instead of the infrared fixed and
quasi--fixed lines. For each fixed set of values of the coupling
constants $Y_t(t_0)$ and $Y_\varkappa(t_0)$, an upper limit on
$Y_\lambda(t_0)$ can be obtained from the requirement that
perturbation theory be applicable up to the Grand Unification
scale $M_X$. A change in the values of the Yukawa coupling
constants $h_t$ and $\varkappa$ at the electroweak scale leads to
a growth or a reduction of the upper limit on $Y_\lambda(t_0)$.
The resulting surface in the
$(\rho_t,\rho_\varkappa,\rho_\lambda)$ space is shown in Figs. 5a
and 5b. In the regime of strong Yukawa coupling, solutions to the
renormalization group equations are concentrated near this
surface. In just the same way as in the case of $Y_\varkappa=0$, a
specific linear combination of $Y_t$, $Y_\lambda$, and
$Y_\varkappa$ is virtually independent of the initial conditions
for $Y_i(0)\to\infty$:
\begin{equation}
\rho_t(t_0)+0.72\rho_{\lambda}(t_0)+0.33\rho_{\varkappa}(t_0)=0.98.
\label{14}
\end{equation}
For $2\le h_t^2(0),\varkappa^2(0),\lambda^2(0)\le 10$, this
combination of the coupling constants,
$h_t^2(t_0)+0.72\lambda^2(t_0)+0.33\varkappa^2(t_0)$, ranges
between $1.35$ and $1.40$; at the same time, we have $1.058\le
h_t^2(t_0)\le 1.219$, $0.032\le \varkappa^2(t_0)\le 0.296$,
$0.098\le\lambda^2(t_0)\le 0.401$ (see Table). The evolution of
$\rho_t(t)+0.72\rho_\lambda(t)+0.33\rho_\varkappa(t)$ at various
initial values of the Yukawa coupling constants is illustrated in
Fig. 6.

On the Hill surface, the region that is depicted in Fig. 5 and
near which the solutions in question are grouped shrinks in one
direction with increasing initial values of the Yukawa coupling
constants, with the result that, at $Y_t(0)$, $Y_\varkappa(0)$,
and $Y_\lambda(0)\sim 1$, all solutions are grouped around the
line that appears as the result of intersection of the
quasi--fixed surface and the infrared fixed surface, which
includes the invariant lines lying in the $\rho_\varkappa=0$ and
$\rho_\lambda=0$ planes and connecting the stable infrared point
with, respectively, the fixed point $\rho_\lambda/\rho_t=1$ and
the fixed point $\rho_\varkappa/\rho_t=1$ in the regime of strong
Yukawa coupling. In the limit $\rho_t, \rho_\lambda,
\rho_\varkappa\gg 1$, in which case the gauge coupling constants
can be disregarded, the fixed points
$\rho_\lambda/\rho_t=1,~\rho_\varkappa/\rho_t=0$ and
$\rho_\varkappa\rho_t=1,~\rho_\lambda/\rho_t=0$ cease to be
stable. Instead of them, the stable fixed point
$R_\lambda=3/4,~R_\varkappa=3/8$ \cite{35} appears in the
$(R_\lambda,R_\varkappa)$ plane, where
$R_\lambda=\rho_\lambda/\rho_t$ and
$R_\varkappa=\rho_\varkappa/\rho_t$. In order to investigate the
behaviour of the solutions to the renormalization group equations
within the NMSSM, it is necessary to linearise the set of these
equations in its vicinity and set $\alpha_i=0$. As a result, we
obtain
\begin{equation}
\begin{split}
R_{\lambda}(t)=&\frac{3}{4}+\left(\frac{1}{2}R_{\lambda 0}+
\frac{1}{\sqrt{5}}R_{\varkappa 0}-\frac{3(\sqrt{5}+1)}{8\sqrt{5}}
\right)\left(\frac{\rho_t(t)}{\rho_{t0}}\right)^{\lambda_1} \\
&{}+\left(\frac{1}{2}R_{\lambda 0}-\frac{1}{\sqrt{5}} R_{\varkappa
0}-\frac{3(\sqrt{5}-1)}{8\sqrt{5}}\right)
\left(\frac{\rho_t(t)}{\rho_{t0}}\right)^{\lambda_2}, \\
R_{\varkappa}(t)=&\frac{3}{8}+\frac{\sqrt{5}}{2}
\left(\frac{1}{2}R_{\lambda 0}+\frac{1}{\sqrt{5}} R_{\varkappa
0}-\frac{3(\sqrt{5}+1)}{8\sqrt{5}}\right)
\left(\frac{\rho_t(t)}{\rho_{t0}}\right)^{\lambda_1} \\
&{}-\frac{\sqrt{5}}{2}\left(\frac{1}{2}R_{\lambda 0}-
\frac{1}{\sqrt{5}}R_{\varkappa 0}-\frac{3(\sqrt{5}-1)}{8\sqrt{5}}
\right)\left(\frac{\rho_t(t)}{\rho_{t0}}\right)^{\lambda_2},\\
\end{split}
\label{15}
\end{equation}
where $R_{\lambda 0}=R_{\lambda}(0)$, $R_{\varkappa
0}=R_{\varkappa}(0)$, $\rho_{t0}=\rho_t(0)$,
$\lambda_1=\dfrac{3+\sqrt{5}}{9}$,
$\lambda_2=\dfrac{3-\sqrt{5}}{9}$, and
$\rho_t(t)=\dfrac{\rho_{t0}}{1+7\rho_{t0}t}$. From (\ref{15}), it
follows that the fixed point $R_\lambda=3/4,~R_\varkappa=3/8$
arises as the result of intersection of two fixed lines in the
$(R_\lambda,R_\varkappa)$ plane. The solutions are attracted most
strongly to the line
$\dfrac{1}{2}R_\lambda+\dfrac{1}{\sqrt{5}}R_\varkappa=\dfrac{3}{8}\left(1+\dfrac{1}{\sqrt{5}}\right)$,
since $\lambda_1\gg \lambda_2$. This line passes through three
fixed points in the $(R_\lambda,R_\varkappa)$ plane: $(1,0)$,
$(3/4,3/8)$, and $(0,1)$. In the regime of strong Yukawa coupling,
the fixed line that corresponds, in the
$(\rho_t,\rho_\varkappa,\rho_\lambda)$ space, to the line
mentioned immediately above is that which lies on the invariant
surface containing a stable infrared fixed point. The line of
intersection of the Hill and the invariant surface can be obtained
by mapping this fixed line into the quasi--fixed surface with the
aid of the set of renormalization group equations. For the
boundary conditions, one must than use the values $\lambda^2(0)$,
$\varkappa^2(0)$, and $h_t^2(0)\gg 1$ belonging to the
aforementioned fixed line.

In just the same way as infrared fixed lines, the infrared fixed
surface is invariant under renormalization group transformations.
In the evolution process, solutions to the set of renormalization
group equations within the NMSSM are attracted to this surface. If
boundary conditions are specified n the fixed surface, the ensuing
evolution of the coupling constants proceeds within this surface.
To add further details, we not that, near the surface being
studied and on it, the solutions are attracted to the invariant
line connecting the stable fixed point
$(\rho_\lambda/\rho_t=3/4,~\rho_\varkappa/\rho_t=3/8)$ in the
regime of strong Yukawa coupling with the stable infrared fixed
point within the NMSSM. In the limit
$\rho_t,\rho_\varkappa,\rho_\lambda\gg 1$, the equation for this
line has the form
\begin{equation}
\begin{split}
\rho_{\lambda}&=\frac{3}{4}\rho_t-\frac{176}{417}+
\frac{3}{139}\rho_2-\frac{7} {417}\rho_1, \\
\rho_{\varkappa}&=\frac{3}{8}\rho_t-\frac{56}{417}-
\frac{18}{139}\rho_2-\frac{68}{2085}\rho_1.
\end{split}
\label{16}
\end{equation}
As one approaches the infrared fixed point, the quantities
$\rho_\lambda$ and $\rho_\varkappa$ tend to zero:
$\rho_\lambda\sim(\rho_t-7/18)^{25/14}$ and
$\rho_\varkappa\sim(\rho_t-7/18)^{9/7}$. This line intersects the
quasi--fixed surface at the point
\[
\rho^{\text{QFP}}_t(t_0)=0.82,\quad
\rho^{\text{QFP}}_{\varkappa}(t_0)=0.087,\quad
\rho^{\text{QFP}}_{\lambda}(t_0)=0.178.
\]
Since all solutions are concentrated in the vicinity of this point
for $Y_t(0), Y_\lambda(0), Y_\varkappa(0)\to\infty$, it should be
considered as a quasi--fixed point for the set of renormalization
group equations within the NMSSM. We note, however, that the
solutions are attracted to the invariant line (\ref{16}) and to
the quasi--fixed line on the Hill surface. This conclusion can be
drawn from the an analysis of the behaviour of the solutions near
the fixed point $(R_\lambda=3/4,~R_\varkappa=3/8)$ (see
Eq.(\ref{15})). Once the solutions have approached the invariant
line
$\dfrac{1}{2}R_\lambda+\dfrac{1}{\sqrt{5}}R_\varkappa=\dfrac{3}{8}\left(1+\dfrac{1}{\sqrt{5}}\right)$,
their evolution is governed by the expression
$(\epsilon(t))^{0.085}$, where $\epsilon(t)=\rho_t(t)/\rho_{t0}$.
This means that the solutions begin to be attracted to the
quasi--fixed point and to the invariant line (\ref{16}) with a
sizable strength only when $Y_i(0)$ reaches a value of $10^2$, at
which perturbation theory is obviously inapplicable. Thus, it is
not the infrared quasi--fixed point but the quasi--fixed line on
the Hill surface (see Fig. 5) that, within the NMSSM, plays a key
role in analysing the behaviour of the solutions to the
renormalization group equations in the regime of strong Yukawa
coupling, where all $Y_i(0)$ are much greater than
$\tilde{\alpha}(0)$.

Along with the invariant surface, which was studied in detail
above, at least three infrared fixed surfaces exist in the
$(\rho_t,\rho_\varkappa,\rho_\lambda)$ space. They attract
solutions to the renormalization group equations much more weakly.
Two of these are specified by the equations $\rho_\lambda$ and
$\rho_\varkappa=0$. Yet another infrared fixed surface can be
found by analysing the behaviour of the solutions in question near
the stable infrared fixed point. Integrating the linearised
renormalisation group equations, we obtain
\begin{equation}
\begin{split}
\rho_t(t)=&\frac{7}{18}+\left(\rho'_{t0}-
\frac{7}{33}\rho_{\lambda 0}+\frac{7}{4}\rho_{20}+
\frac{91}{180}\rho_{10}-\frac{7}{18}\right)
\left(\frac{\tilde{\alpha}_{30}}{\tilde{\alpha}_{3}(t)}\right)^{7/9}\\
&{}+\frac{7}{33}\rho_{\lambda}(t)-\frac{7}{4}\rho_{2}(t)-\frac{91}{180}\rho_{1}(t),\\
\rho_{\lambda}(t)=&\rho_{\lambda
0}\left(\frac{\tilde{\alpha}_{30}}{\tilde{\alpha}_3(t)}\right)^{25/18},\\
\rho_{\varkappa}(t)=&\rho_{\varkappa
0}\left(\frac{\tilde{\alpha}_{30}}{\tilde{\alpha}_3(t)}\right),
\end{split}
\label{17}
\end{equation}
where $\rho'_{t0}$, $\rho_{i0}$, and $\tilde{\alpha}_{30}$ are
constants of integration. In the limiting case of
$\rho_1=\rho_2=0$, the equation of a nontrivial invariant surface
is $\rho_t=\dfrac{7}{18}+\dfrac{7}{33}\rho_\lambda$. This surface
contains nontrivial infrared fixed lines that lie in the
$\rho_\lambda=0$ and $\rho_\varkappa=0$ planes and which weakly
attract solutions to the renormalization group equations. The
inclusion of electroweak interactions significantly modifies the
asymptotic behaviour of this surface near the infrared fixed
point. Nonetheless, a solution to the linearised equations
(\ref{17}) does not fix unambiguously an equations for this
surface. Considering that, at $\rho_\lambda=0$, the equation of
the surface being studied must reduce to the equation for the
invariant line (\ref{12}), we find, for $t\to t_c$, that
\begin{equation}
\rho_t=\frac{7}{18}+\frac{7}{33}\rho_{\lambda}-
6t_c\left(\frac{7}{4}\tilde{\alpha}_2(t_c)+
\frac{91}{180}\tilde{\alpha}_1(t_c)\right)\rho_{\varkappa}.
\label{18}
\end{equation}
Relation (\ref{18}) between $\rho_t$, $\rho_\lambda$, and
$\rho_\varkappa$ is valid for $\rho_\varkappa\gg\rho_\lambda$. By
analysing the behaviour of the solutions in the vicinity of the
stable infrared point (\ref{17}), it can be shown that the
invariant surface (\ref{18}) plays a secondary role in the NMSSM.

\section{Approximate solutions for the Yukawa couplings}

By way of example, the emergence of quasi--fixed lines and
surfaces within the NMSSM can be traced by considering approximate
solutions to the renormalisation group equations from the
Appendix. Recently, approximate solutions of this type were
studied within the minimal SUSY model for $\tan\beta\gg 1$
\cite{37}, in which case $Y_t\sim Y_b\sim Y_\tau$. In the regime
of strong Yukawa coupling within the nonminimal SUSY model, these
solutions are given by
\begin{eqnarray}
\rho_t(t)&=&\frac{E_t(t)}{\tilde{\alpha}_3(t)\left[
6F_t(t)(6F_t(t)+2R_{\lambda 0}F_{\lambda}(t))\right]^{1/2}}+
O\left(\frac{1}{Y_t(0)}\right)+\dots, \nonumber \\
\rho_{\lambda}(t)&=&\frac{R_{\lambda 0} E_{\lambda}(t)}
{\tilde{\alpha}_3(t)\left(6R_{\lambda 0}F_{\lambda}(t)+
6R_{\varkappa 0}t\right)^{1/3}\left(6R_{\lambda
0}F_{\lambda}(t)\right)^{1/6} \left(6F_t(t)+2R_{\lambda
0}F_{\lambda}(t)\right)^{1/2}} \nonumber \\
&&{}+O\left(\frac{1}{Y_t(0)}\right)+\dots, \label{19} \\
\rho_{\varkappa}(t)&=&\frac{R_{\varkappa 0}}{\tilde{\alpha}_3(t)
\left(6R_{\lambda 0}F_{\lambda}(t)+6R_{\varkappa 0}t\right)}+
O\left(\frac{1}{Y_t(0)}\right)+\dots, \nonumber
\end{eqnarray}
where the expressions for the functions $E_i(t)$ and $F_i(t)$ are
presented in the Appendix. Expressions (\ref{19}) for $\rho_i(t)$
were formally obtained by expanding approximate solutions in a
power series in $1/Y_t(0)$. Each subsequent term in such an
expansion is always much less than the preceding one because, in
the approximate solutions, the Yukawa coupling constant for the
$t$--quark always appears in the form of the combination
$Y_t(0)F_t(t)$, which, in the regime of strong Yukawa coupling,
leads to values $\dfrac{1}{Y_t(0)F_t(t)}\ll 1$ at $t\sim t_0$.
From relations (\ref{19}), it follows that, to $O(1/Y_t(0))$
terms, solutions to the renormalisation group equations depend
only on the ratios of the Yukawa coupling constants $R_{\lambda
0}$ and $R_{\varkappa 0}$ at the Grand Unification scale. Setting
$t=t_0$, we obtain a surface in the
$(\rho_t,\rho_\varkappa,\rho_\lambda)$ space. This surface is
specified parametrically; that is, $\rho_i=G_i(R_{\lambda 0},
R_{\varkappa 0})$. Deviations from it are determined by
$O(1/Y_t(0))$ terms, which are negative and small in magnitude in
the limit of strong Yukawa coupling.

However, the approximate solutions (\ref{19}) poorly describe the
evolution of $\rho_\varkappa(t)$.By way of example, we indicate
that, at the electroweak scale, the relative error is about $20\%$
at $\varkappa^2(t_0)\sim 0.1$. This is due above all to the fact
that the self-interaction constant for the scalar field $Y$ is not
renormalised by gauge interactions. The grater the contribution of
gauge interactions to the renormalisation of Yukawa coupling
constants, the higher the accuracy to which the approximate
solutions describe their evolution. For example, the relative
error in $\rho_t(t_0)$ ($\rho_\lambda(t_0)$) is $2-3\%$ (about
$5-6\%$) at $Y_t(0)\sim Y_\varkappa(0)\sim Y_\lambda(0)$.

An approximate solution for $Y_\varkappa=0$ and
$Y_t(0),Y_\lambda(0)\gg\tilde{\alpha}_i(0)$ can be obtained by
setting $R_{\varkappa 0}=0$ in Eqs.(\ref{19}). In the regime of
strong Yukawa coupling, $\rho_t(t)$ and $\rho_\lambda(t)$ then
depend only on $R_{\lambda 0}$, with the result that, in the
$(\rho_t,\rho_\lambda)$ plane, there arises, at $t=t_0$, the Hill
line
\begin{equation}
\rho^2_t+\frac{1}{3}\left(\frac{E_t(t_0)}{E_{\lambda}(t_0)}\right)^2
\left(\frac{F_{\lambda}(t_0)}{F_t(t_0)}\right)^2\rho^2_{\lambda}=
\rho^2_{\text{QFP}}, \label{20}
\end{equation}
where
$\rho_{\text{QFP}}=\dfrac{E_t(t_0)}{6F_t(t_0)\alpha_3(t_0)}$. With
increasing initial values of the Yukawa coupling constants,
$O(1/Y_t(0))$ terms, which determine the deviation of the
solutions in question from the quasi--fixed line (\ref{20}),
decrease, so that the approximate solutions to the renormalisation
group equations within the NMSSM are attracted to this line. The
explicit form of the dependences $\rho_\lambda(t)$ and $\rho_t(t)$
in (\ref{19}) makes it possible to find that the invariant line
lying in the $(\rho_t,\rho_\lambda)$ plane and corresponding to
$R_{\lambda 0}=1$ can be approximately parametrised as
\begin{equation}
\begin{split}
\rho_t(t)&=\frac{E_t(t)}{\tilde{\alpha}_3(t)\left[
6F_t(t)(6F_t(t)+2F_{\lambda}(t))\right]^{1/2}}\\
\rho_{\lambda}(t)&=\frac{E_{\lambda}(t)}{\tilde{\alpha}_3(t)\left[6
F_{\lambda}(t)
\left(6F_t(t)+2F_{\lambda}(t)\right)\right]^{1/2}}\, .
\end{split}
\label{21}
\end{equation}
The values $\rho_t(t_0)$ and $\rho_\lambda(t_0)$ as calculated by
formulas (\ref{21}) are the coordinates of the point where the
Hill fixed line (\ref{20}) intersects the infrared fixed line
(\ref{21}), which appears to be a quasi--fixed point for the set
of renormalisation group equations within the NMSSM. Our numerical
results, which are displayed in Fig. 7, demonstrate that relations
(\ref{20}) and (\ref{21}) reproduce quite accurately the
quasi--fixed and the invariant line at $R_{\lambda 0}\le 1$.
Significant deviations are observed only in the infrared region
($t\to t_c$) and for $R_{\lambda 0}\gg 1$. In general, the
relative deviation of the approximate solution in question from
the exact one is $5-6\%$ at $\rho_\varkappa=0$ and $R_{\lambda
0}\sim 1$ and grows fast with increasing $\rho_\lambda/\rho_t$.

\section{Unification of the Yukawa couplings $h_b$ and $h_\tau$}

As was indicated above, Grand Unification Theory imposes
additional constraints on the parameter space of SUSY models.
Among such constraints, the unification of the Yukawa coupling
constants for the $b$--quark and the $\tau$--lepton at the scale
$M_X$ is worthy of note above all. In the nonminimal SUSY model,
$h_b$ and $h_\tau$ are unified if the constants $Y_t$,
$Y_\lambda$, and $Y_\varkappa$ satisfy the relations:
\begin{equation}
\begin{split}
\frac{Y_t(0)}{Y_t(t_0)}&=\left[\frac{R_{b\tau}(0)}{R_{b\tau}(t_0)}
\right]^{\sfrac{21}{2}}
\left[\frac{\tilde{\alpha}_3(t_0)}{\tilde{\alpha}_3(0)}\right]^{\sfrac{68}{9}}
\left[\frac{\tilde{\alpha}_2(t_0)}{\tilde{\alpha}_2(0)}\right]^{\sfrac{9}{4}}
\left[\frac{\tilde{\alpha}_1(t_0)}{\tilde{\alpha}_1(0)}\right]^{\sfrac{463}{396}}
\left[\frac{Y_{\lambda}(0)}{Y_{\lambda}(t_0)}\right]^{\sfrac{1}{4}},\\
\frac{Y_t(0)}{Y_t(t_0)}&=\left[\frac{R_{b\tau}(0)}{R_{b\tau}(t_0)}
\right]^{9}
\left[\frac{\tilde{\alpha}_3(t_0)}{\tilde{\alpha}_3(0)}\right]^{\sfrac{56}{9}}
\left[\frac{\tilde{\alpha}_2(t_0)}{\tilde{\alpha}_2(0)}\right]^{\sfrac{3}{2}}
\left[\frac{\tilde{\alpha}_1(t_0)}{\tilde{\alpha}_1(0)}\right]^{\sfrac{197}{198}}
\left[\frac{Y_{\lambda}(0)}{Y_{\lambda}(t_0)}\right]^{\sfrac{1}{2}}
\left[\frac{Y_\varkappa(t_0)}{Y_\varkappa(0)}\right]^{\sfrac{1}{6}},
\end{split}
\label{22}
\end{equation}
where $R_{b\tau}(t_0)=m_b(t_0)/m_\tau(t_0)$ is the ratio of the
running masses of the $b$--quark and the $\tau$--lepton at the
electroweak scale; in the minimal unification schemes we have
$R_{b\tau}(0)=\sqrt{\dfrac{Y_b(0)}{Y_\tau(0)}}=1$. The equation
determining the function $R_{b\tau}(t)$ is presented in the
Appendix (see Eq.(\ref{A.2})). The first relation in (\ref{22})
corresponds to the case of $\varkappa=0$, whereas the second
implies a $\varkappa$ value different from zero. Relations
(\ref{22}) can be obtained by directly integrating the
renormalisation group equations. Setting $R_{b\tau}(t_0)=1.61$,
which corresponds to $m_b(t_0)=2.86\text{~GeV}$ and
$m_\tau(t_0)=1.78\text{~GeV}$, we find, for the ratio of the
Yukawa coupling constants for the $t$--quark, that
\begin{equation}
\frac{Y_t(0)}{Y_t(t_0)}\approx 3.67
\left[\frac{Y_{\lambda}(0)}{Y_{\lambda}(t_0)}\right]^{1/4},\qquad
\frac{Y_t(0)}{Y_t(t_0)}\approx 2.57
\left[\frac{Y_{\lambda}(0)}{Y_{\lambda}(t_0)}\right]^{1/2}
\left[\frac{Y_\varkappa(t_0)}{Y_\varkappa(0)}\right]^{1/6}.
\label{23}
\end{equation}
The second equation in (\ref{23}) -- it relates $Y_t$,
$Y_\lambda$, and $Y_\varkappa$ -- determines a surface in the
$(\rho_t,\rho_\lambda,\rho_\varkappa)$ space; at $Y_\varkappa=0$,
this surface degenerates into a line in the
$(\rho_t,\rho_\lambda)$ plane. In this case, $b-\tau$ unification
is possible under the condition $Y_t(0)\gg Y_t(t_0)$, which is
realised only in the regime of strong Yukawa coupling within the
NMSSM. In the $(\rho_t,\rho_\lambda)$ plane, Fig. 8 shows the Hill
line and the curve that corresponds to $Y_b(0)=Y_\tau(0)$. As
might have been expected, the spacing between them is quite small.
In addition, we note that only at sufficiently large values of the
$t$--quark Yukawa coupling constant at the electroweak scale,
$Y_t(t_0)>Y_t^0$, is $b-\tau$ unification possible. A lower limit
on the $Y_t(t_0)$ implies that there exists an upper limit on
$\tan\beta$ (see Eq.(\ref{1})). By varying the running $b$--quark
mass $m_b(m_b)$ from 4.1 to 4.4 GeV, we found that only for
$\tan\beta\le 2$ can the equality of the Yukawa coupling constants
at the Grand Unification scale be achieved. The possibility of
unifying the Yukawa coupling constants within the NMSSM was
investigated in detail elsewhere \cite{38}. The condition
$Y_b(0)=Y_\tau(0)$ imposes stringent constraints on the parameter
space of the model being studied. Since $h_b$ and $h_\tau$ are
small in magnitude at $\tan\beta\sim 1$, they can be generated,
however, at the Grand Unification scale owing to unrenormalised
operators upon a spontaneous breakdown of symmetry, in which case
$h_b$ and $h_\tau$ can take different values.

\section{Conclusion}

The present analysis has revealed that, in the regime of strong
Yukawa coupling, solutions to the renormalisation group equations
within the NMSSM, $Y_i(t)$, are attracted to quasi--fixed lines
and surfaces in the space of Yukawa coupling constants and that
specific combinations $\rho_i(t)$ are virtually independent of
their initial values at the Grand Unification scale. It is for
$Y_i(0)\gg\tilde{\alpha}_i(0)$ that the upper limit on the mass of
the lightest Higgs boson attains its maximum value. It has also
been proven that, in the limit being considered, the values of the
constants $h_t$, $\lambda$, and $\varkappa$ can be chosen in such
a way as to ensure unification of the Yukawa coupling constants
for the $b$--quark and $\tau$--lepton at the scale $M_X$, a
feature usually inherent in Grand Unified Theories. For
$Y_i(0)\to\infty$, all solution to the renormalisation group
equations are concentrated near quasi--fixed points. These points
emerge as the result of intersection of Hill lines or surfaces
with the invariant line that connects the stable fixed point for
$Y_i\gg\tilde{\alpha}_i$ with the stable infrared fixed point. For
the renormalisation group equations within the NMSSM, we have
listed all the most important invariant lines and surfaces and
studied their asymptotic behaviour for $Y_i\gg\tilde{\alpha}_i$
and in the vicinity of the infrared fixed point.

With increasing $Y_i(0)$, the solutions in question approach
quasi--fixed points quite slowly; that is, the deviation is
proportional to $(\epsilon_t(t))^\delta$, where
$\epsilon_t(t)=Y_t(t)/Y_t(0)$ and $\delta$ is calculated by
analysing the set of the renormalisation group equations in the
regime of strong Yukawa coupling. As a rule, $\delta$ is positive
and much less than unity. By way of example, we indicate that, in
the case where all three Yukawa coupling constants differ from
zero, $\delta\approx 0.085$. Of greatest importance in analysing
the behaviour of solutions to the renormalisation group equations
within the NMSSM at $Y_t(0),Y_\lambda(0),Y_\varkappa(0)\sim 1$ is
therefore not the infrared quasi--fixed point but the line lying
on the Hill surface and emerging as the intersection of the Hill
and invariant surface. This line can be obtained by mapping the
fixed points $(1,0)$, $(3/4,3/8)$, and $(0,1)$ in the
$(R_\lambda,R_\varkappa)$ plane for $Y_i\gg\tilde{\alpha}_i$ into
the quasi--fixed surface by means of renormalisation group
equations.

The emergence of Hill lines and surfaces in the space of Yukawa
coupling constants can be traced by considering the examples of
approximate solutions that are presented in the Appendix. These
solutions lead to qualitatively correct results. However, the
approximate solutions poorly describe the evolution of
$Y_\varkappa(t)$, since the neutral field $Y$ is not renormalised
by gauge interactions. At the same time, it has been shown that,
at $Y_t(0)\sim Y_\lambda(0)$, the relative deviation of the
approximate solution from the exact one is as small as $2-3\%$ in
$Y_t(t_0)$ and about $5-6\%$ in $Y_\lambda(t_0)$. With increasing
$Y_\lambda(t_0)/Y_t(t_0)$, such relative deviations grow quite
fast.

\section*{Acknowledgements}

The authors are grateful to M.I.Vysotsky, D.I.Kazakov, and
K.A.Ter-Martirosyan for stimulating questions, valuable
discussions, and enlightening comments. R.B.Nevzorov is indebted
to Insituto Nazionale di Fisica Nucleare (Sezione di Ferrara) for
the kind hospitality extended to him.

This work was supported by the Russian Foundation for Basic
Research (RFBR), projects \#\# 98-02-17372, 98-02-17453,
96-15-96578, 96-15-96740; by INTAS, project \# 93-3316-Ext.; by a
joint RFBR--INTAS grant \# 95-0567.

\newpage

\section*{Appendix}

{\bfseries Set of the renormalisation group equations of the NMSSM
for the Yukawa couplings and an approximate solution to it.}

In the present study we have analysed the one--loop
renormalisation group equations within the NMSSM. These equations
can be represented as \cite{34}:
\begin{eqnarray}
\frac{d\tilde{\alpha}_i}{dt}&=&-b_i\tilde{\alpha}_i^2,\nonumber \\
\frac{dY_t}{dt}&=&-Y_t(Y_{\lambda}+6Y_t
-\frac{16}{3}\tilde{\alpha}_3-3\tilde{\alpha}_2-
\frac{13}{15}\tilde{\alpha}_1),\label{A.1} \\
\frac{dY_{\lambda}}{dt}&=&-Y_{\lambda}(4Y_{\lambda}+2Y_{\varkappa}+3Y_t
 -3\tilde{\alpha}_2-\frac{3}{5}\tilde{\alpha}_1),\nonumber \\
\frac{dY_\varkappa}{dt}&=&-6Y_{\varkappa}(Y_{\lambda}+Y_{\varkappa}).
\nonumber
\end{eqnarray}
On the right--hand sides of these differential equations we have
discarded terms proportional to the Yukawa couplings $Y_b$ and
$Y_\tau$, since their contribution at $\tan\beta\ll 10$ is
negligibly small. The index $i$ runs through the values from 1 and
3, $b_1=\dfrac{33}{5}$, $b_2=1$, $b_3=-3$,
$\tilde{\alpha}_i(t)=\dfrac{\alpha_i(t)}{4\pi}=\left(\dfrac{g_i(t)}{4\pi}\right)^2$,
$Y_t(t)=\left(\dfrac{h_t(t)}{4\pi}\right)^2$,
$Y_\lambda(t)=\left(\dfrac{\lambda(t)}{4\pi}\right)^2$, and
$Y_\varkappa(t)=\left(\dfrac{\varkappa(t)}{4\pi}\right)^2$. The
variable $t$ is defined in a standard way:
$t=\ln\left(\dfrac{M_X^2}{Q^2}\right)$, where $M_X=3\cdot
10^{16}\text{~GeV}$.

In analysing $b-\tau$ unification, use was made of the evolution
equation for the ratio
$R_{b\tau}(t)=\sqrt{\dfrac{Y_b(t)}{Y_\tau(t)}}$ of the Yukawa
couplings for the $b$--quark and the $\tau$--lepton,
\begin{equation}
\label{A.2} \frac{dR_{b\tau}}{dt}=-R_{b\tau}
\left(\frac{1}{2}Y_t-\frac{8}{3}\tilde{\alpha}_3
+\frac{2}{3}\tilde{\alpha}_1\right),
\end{equation}
where $Y_b=\left(\dfrac{h_b(t)}{4\pi}\right)^2$ and
$Y_\tau(t)=\left(\dfrac{h_\tau(t)}{4\pi}\right)^2$. The value of
$R_{b\tau}(0)=1$ corresponds to the unification of the Yukawa
couplings $h_b$ and $h_\tau$. For the Yukawa and gauge couplings
the set of two--loop renormalisation group equations within the
NMSSM is presented in \cite{30}.

The corresponding one--loop equations for the gauge couplings can
easily be integrated. The result has the form:
\begin{equation}
\label{A.3}
\tilde{\alpha}_i(t)=\frac{\tilde{\alpha}_i(0)}{1+b_i\tilde{\alpha}_i(0)t}.
\end{equation}
Since the gauge couplings within the MSSM and within the NMSSM
coincide at the Grand Unification scale, we have
$\tilde{\alpha}_i(0)=\tilde{\alpha}(0)=\tilde{\alpha}_{\text{GUT}}$
for all of them. In the case where $\lambda=0$, there exists an
exact analytic solution to the set of renormalisation group
equations (\ref{A.1}). It is specified by relations (\ref{2}) and
(\ref{8}), with $E(t)$ and $F(t)$ being given by
\begin{eqnarray}
H(t)&=&\frac{16}{3}\tilde{\alpha}_3(t)+3\tilde{\alpha}_2(t)
 +\frac{13}{15}\tilde{\alpha}_1(t),\nonumber \\
E(t)&=&\exp\left[\int\limits_0^t H(t')dt'\right]=
\left[\frac{\tilde{\alpha}_3(t)}{\tilde{\alpha}(0)}\right]^{16/9}
\left[\frac{\tilde{\alpha}_2(t)}{\tilde{\alpha}(0)}\right]^{-3}
\left[\frac{\tilde{\alpha}_1(t)}{\tilde{\alpha}(0)}\right]^{-13/99},\label{A.4}
\\ F(t)&=&\int\limits_0^tE(t')dt'. \nonumber
\end{eqnarray}

In the regime of strong Yukawa coupling, in which case all
$Y_i(0)$ are much greater than $\tilde{\alpha}(0)$, an exact
analytic solution to the set of Eqs.(\ref{A.1}) has not yet been
found. An explicit $t$ dependence of the Yukawa couplings can be
obtained on the basis of an approximate solution to the
renormalisation group equations within the NMSSM. An approximate
solution of this type was first obtained by Kazakov \cite{37}, who
studied the renormalisation group equations within the MSSM in the
limit $\tan\beta\gg 1$. For the Yukawa couplings, it has the form:
\begin{eqnarray}
Y_{\lambda}(t)&=&Y_{\lambda}(0)E_{\lambda}(t)P_2(t)P_1(t)P_Y(t),\nonumber
\\ Y_{\varkappa}(t)&=&Y_{\varkappa}(0)P_Y^3(t),\label{A.5} \\
Y_t(t)&=&Y_t(0)E_{t}(t)P_Q(t)P_U(t)P_2(t),\nonumber
\end{eqnarray}
where
\[ E_t(t)=E(t),\qquad E_{\lambda}(t)=
\left(\frac{\tilde{\alpha}_2(t)}{\tilde{\alpha}(0)}\right)^{-3}
\left(\frac{\tilde{\alpha}_1(t)}{\tilde{\alpha}(0)}\right)^{-1/11},
\]
and $P_i(t)$ is the contribution of the Yukawa couplings to the
renormalisation of $Y_i(t)$ from each of the external legs
entering the corresponding vertex:
\begin{equation}
\label{A.6}
\begin{gathered}
\frac{d\ln P_Q(t)}{dt}=\frac{1}{2}\frac{d\ln
P_U(t)}{dt}=-Y_t(t),\qquad \frac{d\ln
P_2(t)}{dt}=-3Y_t(t)-Y_{\lambda}(t),\\ \frac{d\ln
P_1(t)}{dt}=-Y_{\lambda}(t),\qquad \frac{d\ln
P_Y(t)}{dt}=-2Y_{\lambda}(t)-2Y_{\varkappa}(t).
\end{gathered}
\end{equation}
Setting $P_Q(t)P_U(t)P_2(t)\approx P_2(t)P_1(t)P_Y(t)\approx
P_Y^3(t)\approx P_0(t)$ and $P_Q^A(t)\approx P_U^B(t)\approx
P_2^{C_2}(t)\approx P_1^{C_1}(t)\approx P_Y^D(t)\approx P_0(t)$,
we find that $A$, $B$, $C_1$, $C_2$, and $D$ satisfy the
relations:
\[ \frac{1}{A}+\frac{1}{B}+\frac{1}{C_2}=1,\quad \frac{1}{D}+\frac{1}{C_1}+\frac{1}{C_2}=1,\quad D=3. \]
Since the contribution of the $t$--quark Yukawa coupling to the
renormalisation of external legs is much greater than the
contribution of $Y_\lambda$, the constants $A$, $B$, and $C_2$
also satisfy the approximate relations $B\approx A/2$ and
$C_2\approx A/3$, which make it possible to find, for $A$, $B$,
$C_1$, and $C_2$, that
\[ A=C_1=6,\quad B=3,\quad C_2=2 \]
and to obtain, with the aid of the differential equations
(\ref{A.6}) for $P_i(t)$, approximate solutions for the latter
quantities. The results are
\begin{eqnarray}
P_U(t)&=&\frac{1}{(1+6Y_t(0)F_t(t))^{1/3}}=P_Q^2(t),\nonumber \\
P_{H_2}(t)&=&\frac{1}{(1+6Y_t(0)F_t(t)+2Y_{\lambda}(0)
F_{\lambda}(t))^{1/2}},\label{A.7} \\
P_{H_1}(t)&=&\frac{1}{(1+6Y_{\lambda}(0)F_{\lambda}(t))^{1/6}},\nonumber
\\
P_Y(t)&=&\frac{1}{(1+6Y_{\lambda}(0)F_{\lambda}(t)+6Y_{\varkappa}(0)
t)^{1/3}},\nonumber
\end{eqnarray}
where
\[ F_t(t)=F(t),\qquad F_{\lambda}(t)=\int_0^t
E_{\lambda}(t')dt'. \]

Substituting the resulting expressions (\ref{A.7}) for $P_i(t)$
into relations (\ref{A.5}), we obtain approximate solutions for
the Yukawa couplings within the NMSSM. In just the same way, we
can find approximate solutions for $Y_t(t)$ and $Y_\lambda(t)$ at
$\varkappa=0$. As a result, it can easily be proven that the
required solutions are obtained by setting $Y_\varkappa(0)=0$ in
(\ref{A.5}) and (\ref{A.7}).

\newpage

\begin{center}
{\bfseries Table.} Values of the Yukawa couplings at the
electroweak scale for various initial values $\varkappa^2(0)$,
$\lambda^2(0)$, and $h_t^2(0)$.
\end{center}

\hspace*{-2mm}\noindent%
\begin{tabular}{|c|c|c|c|c|c|c|c|}
\hline $\varkappa^2(0)$ & $\lambda^2(0)$ & $h_t^2(0)$ &
$\varkappa^2(t_0)$ & $\lambda^2(t_0)$ & $h_t^2(t_0)$ &
$h_t^2(t_0)+0.51\lambda^2(t_0)$ & $h_t^2(t_0)+0.72\lambda^2(t_0)$
\\ &&&&&&& $+0.33\varkappa^2(t_0)$ \\ \hline
0&10&10&0&0.3220&1.1538&1.3180&1.3857\\ \hline
0&6&10&0&0.2879&1.1675&1.3143&1.3747\\ \hline
0&2&10&0&0.1911&1.1987&1.2962&1.3363\\ \hline
0&10&6&0&0.3492&1.1327&1.3108&1.3841\\ \hline
0&6&6&0&0.3167&1.1475&1.3090&1.3755\\ \hline
0&2&6&0&0.2203&1.1815&1.2939&1.3402\\ \hline
0&10&2&0&0.4209&1.0513&1.2659&1.3543\\ \hline
0&6&2&0&0.3901&1.0715&1.2704&1.3524\\ \hline
0&2&2&0&0.2941&1.1160&1.2660&1.3277\\ \hline
10&10&10&0.1480&0.2480&1.1737&1.3002&1.4011\\ \hline
10&6&10&0.1995&0.1969&1.1904&1.2908&1.3980\\ \hline
10&2&10&0.2956&0.0979&1.2193&1.2692&1.3874\\ \hline
10&10&6&0.1256&0.2801&1.1527&1.2956&1.3958\\ \hline
10&6&6&0.1760&0.2279&1.1712&1.2875&1.3934\\ \hline
10&2&6&0.2785&0.1192&1.2047&1.2655&1.3825\\ \hline
10&10&2&0.0865&0.3601&1.0734&1.2570&1.3612\\ \hline
10&6&2&0.1305&0.3060&1.0984&1.2545&1.3618\\ \hline
10&2&2&0.2385&0.1775&1.1458&1.2363&1.3523\\ \hline
2&10&10&0.0655&0.2941&1.1608&1.3108&1.3942\\ \hline
2&6&10&0.1055&0.2482&1.1767&1.3033&1.3903\\ \hline
2&2&10&0.2059&0.1396&1.2092&1.2804&1.3777\\ \hline
2&10&6&0.0521&0.3244&1.1395&1.3049&1.3903\\ \hline
2&6&6&0.0875&0.2798&1.1567&1.2994&1.3870\\ \hline
2&2&6&0.1865&0.1663&1.1929&1.2778&1.3742\\ \hline
2&10&2&0.0322&0.4007&1.0582&1.2625&1.3573\\ \hline
2&6&2&0.0578&0.3581&1.0810&1.2637&1.3580\\ \hline
2&2&2&0.1464&0.2361&1.1297&1.2501&1.3480\\ \hline
\end{tabular}

\newpage

\section*{Figure captions}

{\bfseries Fig.1.} Boundary conditions for the renormalisation
group equations of the NMSSM at the scale $q=M_X$ for
$\varkappa^2=0$ uniformly distributed in the
$(\rho_t,\rho_\lambda)$ plane in a square $2\le
h_t^2(0),\lambda^2(0)\le 10$ -- Fig.1a, and the corresponding
values of the Yukawa couplings at the electroweak scale -- Fig.1b.
The thick and thin curves in Fig.1b represent, respectively, the
invariant and the Hill line. The dashed line in Fig.1b is a fit of
the values $(\rho_t(t_0),\rho_\lambda(t_0))$ for $20\le
h_t^2(0),\lambda^2(0)\le 100$.\\

{\bfseries Fig.2.} Boundary conditions for the renormalisation
group equations of the NMSSM at the scale $q=M_X$ for
$\varkappa^2=0$ uniformly distributed in the
$(\rho_t,\rho_\lambda)$ plane in a square $20\le
h_t^2(0),\lambda^2(0)\le 100$ -- Fig.2a, and the corresponding
values of the Yukawa couplings at the electroweak scale -- Fig.2b.
The thick and thin curves in Fig.2b represent, respectively, the
invariant and the Hill line. The dashed line in Fig.2b is a fit of
the values $(\rho_t(t_0),\rho_\lambda(t_0))$ for $20\le
h_t^2(0),\lambda^2(0)\le 100$.\\

{\bfseries Fig.3.} Evolution of the combination
$\rho_t(t)+0.506\rho_\lambda(t)$ of the Yukawa couplings from the
Grand Unification scale ($t=0$) to the electroweak scale ($t=t_0$)
for $\varkappa^2=0$ and for various initial values $h_t^2(0)$ --
Fig.3a, $\lambda^2(0)$ -- Fig.3b.\\

{\bfseries Fig.4.} Boundary conditions for the renormalisation
group equations of the NMSSM at the scale $q=M_X$ for
$\lambda^2=0$ uniformly distributed in the
$(\rho_t,\rho_\varkappa)$ plane in a square $2\le
h_t^2(0),\varkappa^2(0)\le 10$ -- Fig.4a, and the corresponding
values of the Yukawa couplings at the electroweak scale -- Fig.4b.
The thick and thin curves in Fig.4b represent, respectively, the
invariant and the Hill line.\\

{\bfseries Fig.5.} Quasi--fixed surface in the
$(\rho_t,\rho_\varkappa,\rho_\lambda)$ space. The shaded part of
the surface represents the region near which the solutions
corresponding to the initial values $2\le
h_t^2(0),\varkappa^2(0),\lambda^2(0)\le 10$ -- Fig.5a, $20\le
h_t^2(0),\varkappa^2(0),\lambda^2(0)\le 100$ -- Fig.5b are
concentrated.\\

{\bfseries Fig.6.} Evolution of the combination
$\rho_t+0.720\rho_\lambda+0.3330\rho_\varkappa$ of the Yukawa
couplings from the Grand Unification scale ($t=0$) to the
electroweak scale ($t=t_0$) for various initial values $h_t^2(0)$
-- Fig.6a, $\lambda^2(0)$ -- Fig.6b, $\varkappa^2(0)$ -- Fig.6c.\\

{\bfseries Fig.7.} Invariant and quasi--fixed lines in the
$(\rho_t,\rho_\lambda)$ plane obtained by means of numerical
calculations (thick and thin solid curves) and by means of
approximate solution of the renormalisation group equations of the
NMSSM (dashed and dotted curves). The thick solid curve and the
dashed curve represent the infrared fixed line, while the thin
solid curve and the dotted curve represent the Hill line.\\

{\bfseries Fig.8.} The values of the Yukawa couplings at the
electroweak scale corresponding to the unification of $h_b$ and
$h_\tau$ at the scale $q=M_X$ (thick line). The infrared
quasi--fixed line is also presented here (thin line).

\newpage

\begin{landscape}
\noindent%
\includegraphics[height=93mm,keepaspectratio=true]{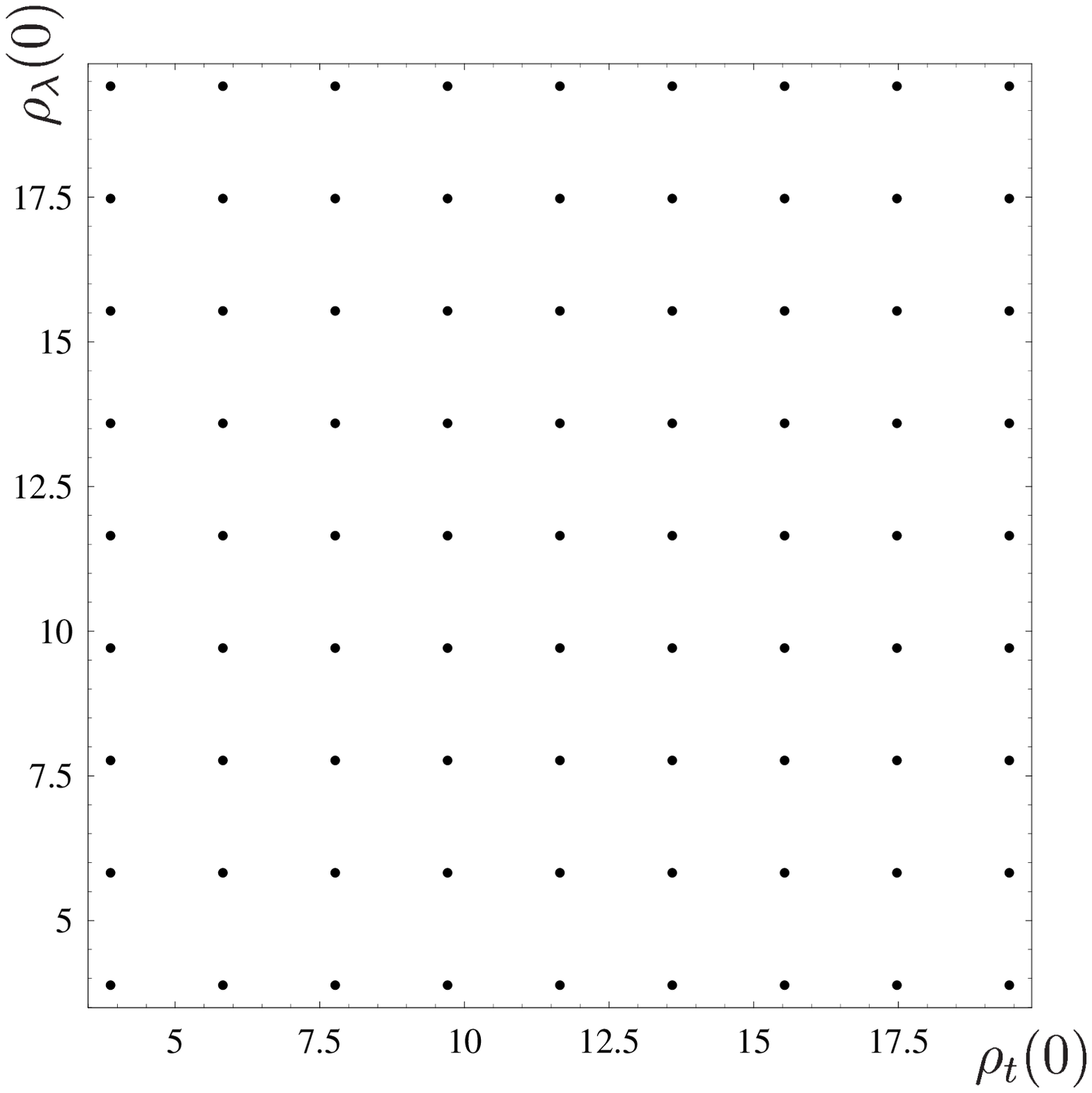}
\hfill\raisebox{-2mm}%
{\includegraphics[height=99mm,keepaspectratio=true]{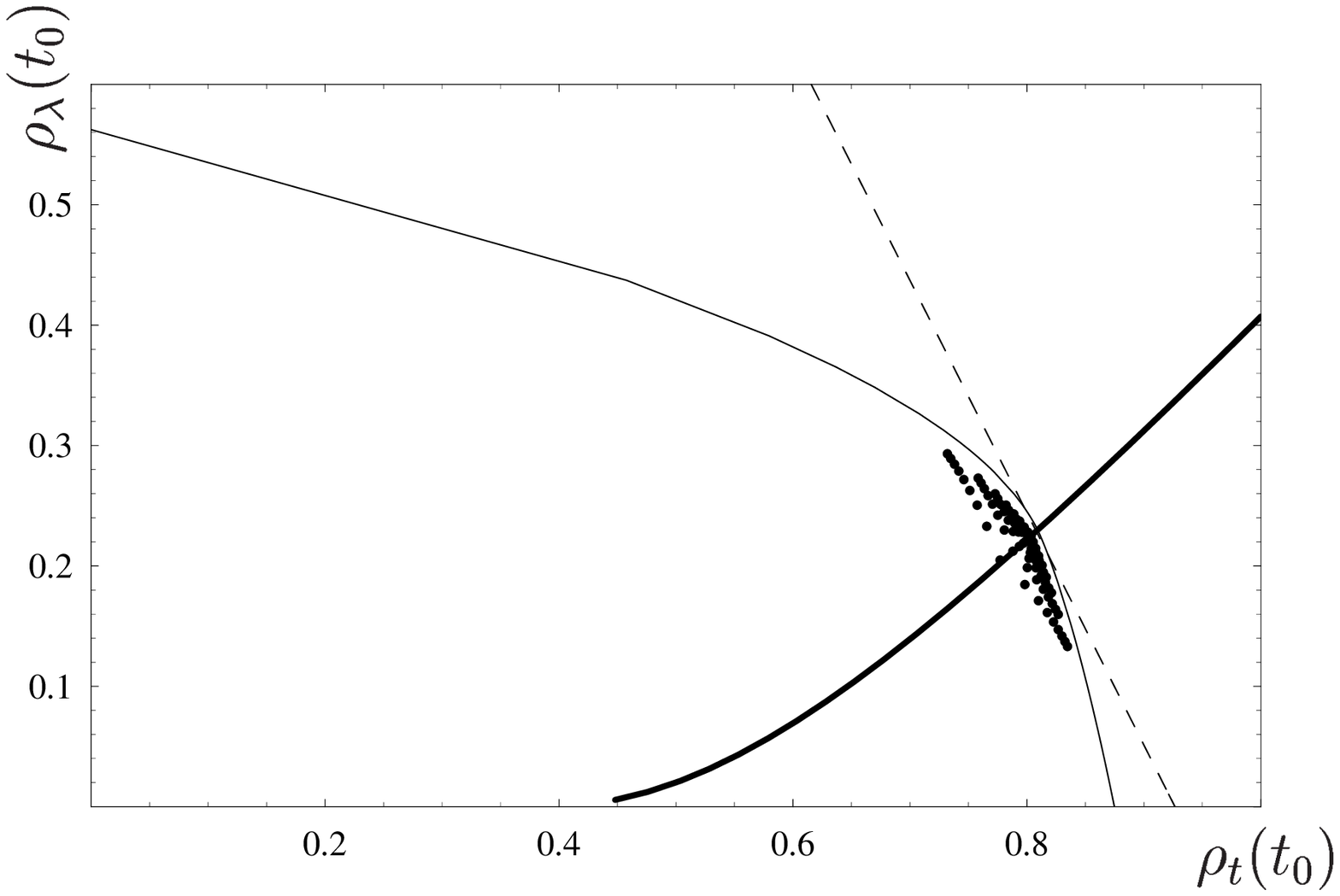}}

\vspace{5mm}\hspace*{32mm}{\large\bfseries Fig.1a.
}\hspace{107mm}{\large\bfseries Fig.1b.}

\end{landscape}

\begin{landscape}
\noindent%
\includegraphics[height=93mm,keepaspectratio=true]{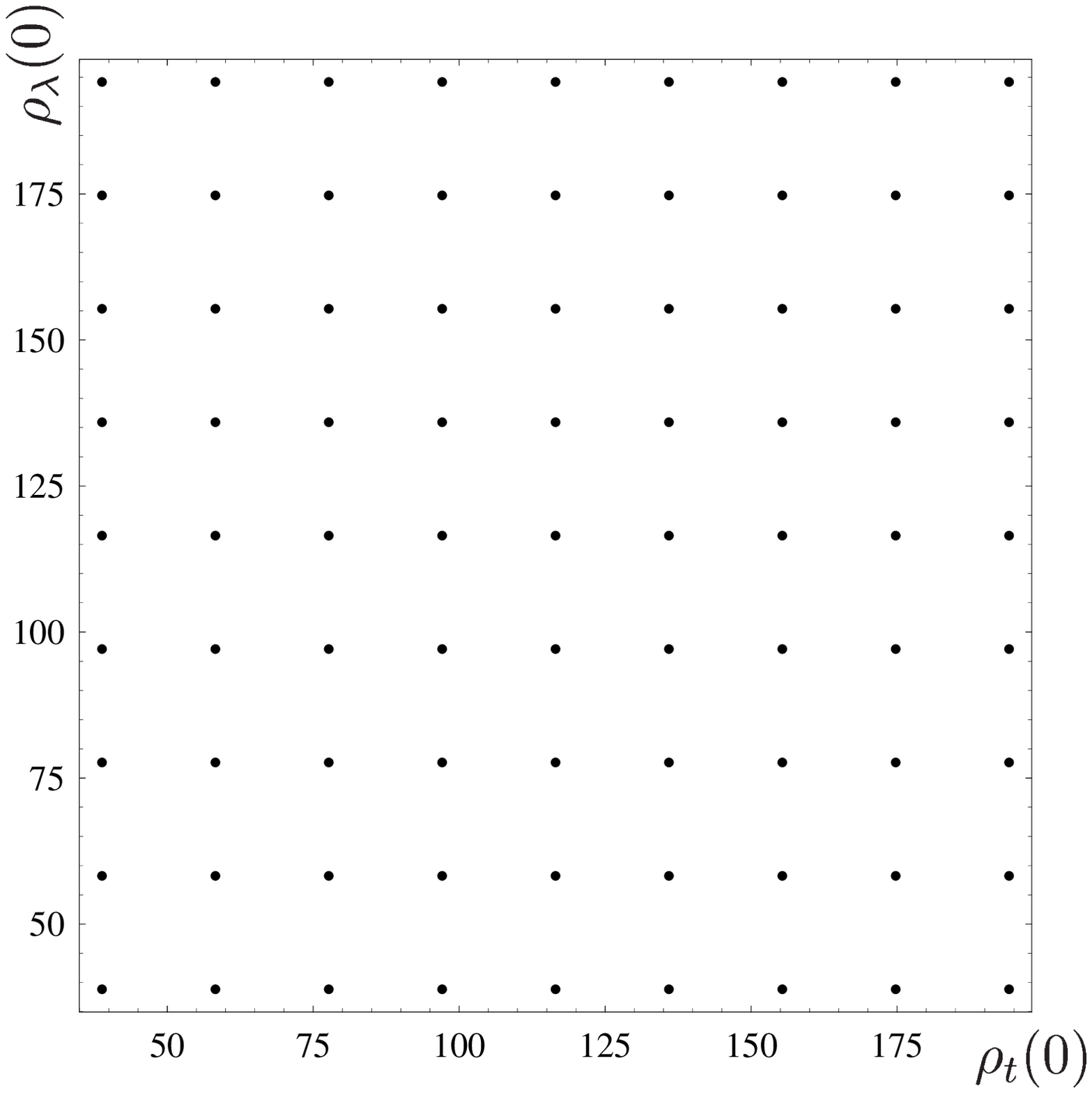}
\hfill\raisebox{-2mm}%
{\includegraphics[height=99mm,keepaspectratio=true]{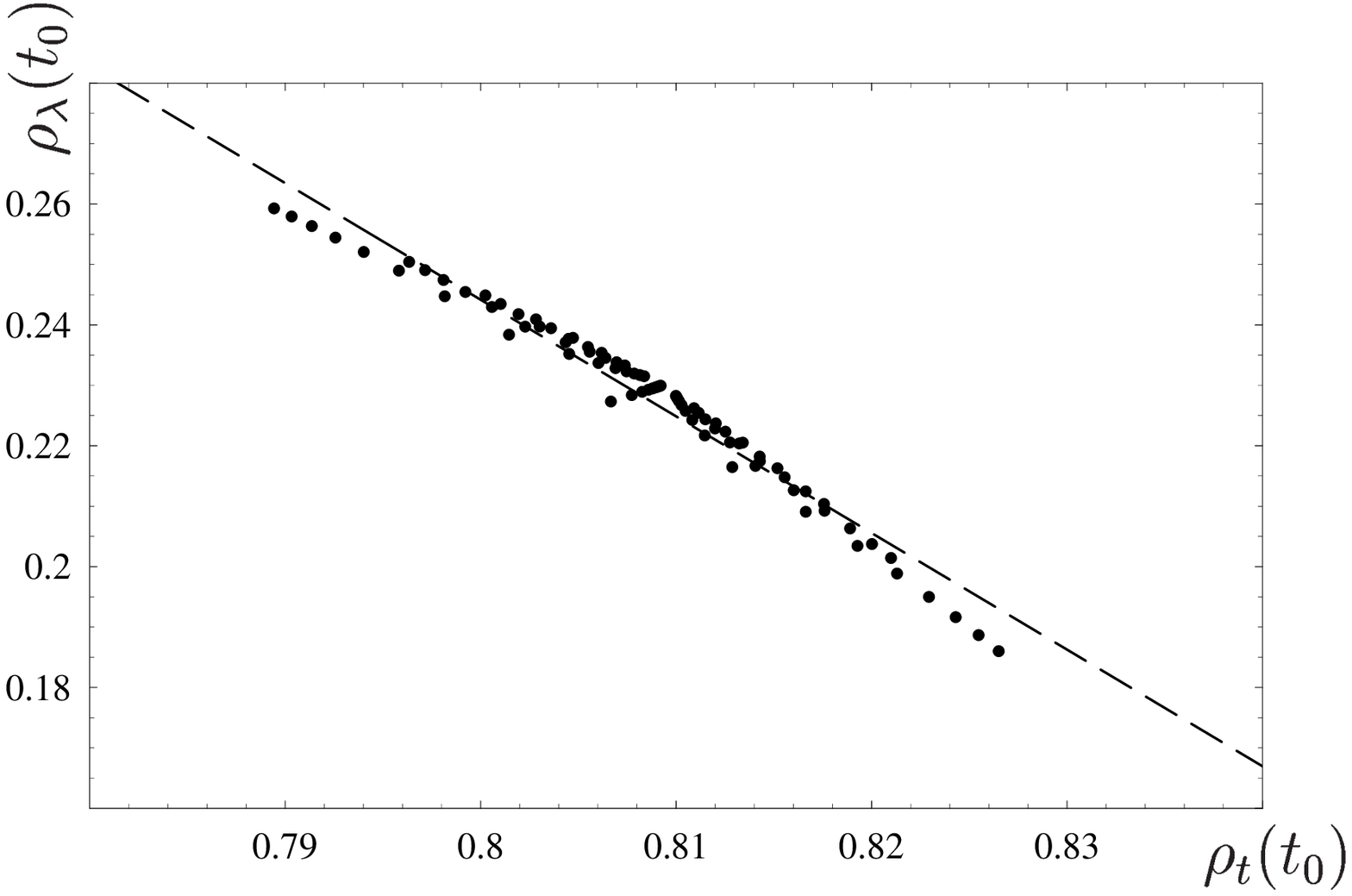}}

\vspace{5mm}\hspace*{32mm}{\large\bfseries Fig.2a.
}\hspace{107mm}{\large\bfseries Fig.2b.}

\end{landscape}

\begin{center}

\includegraphics[height=96mm,keepaspectratio=true]{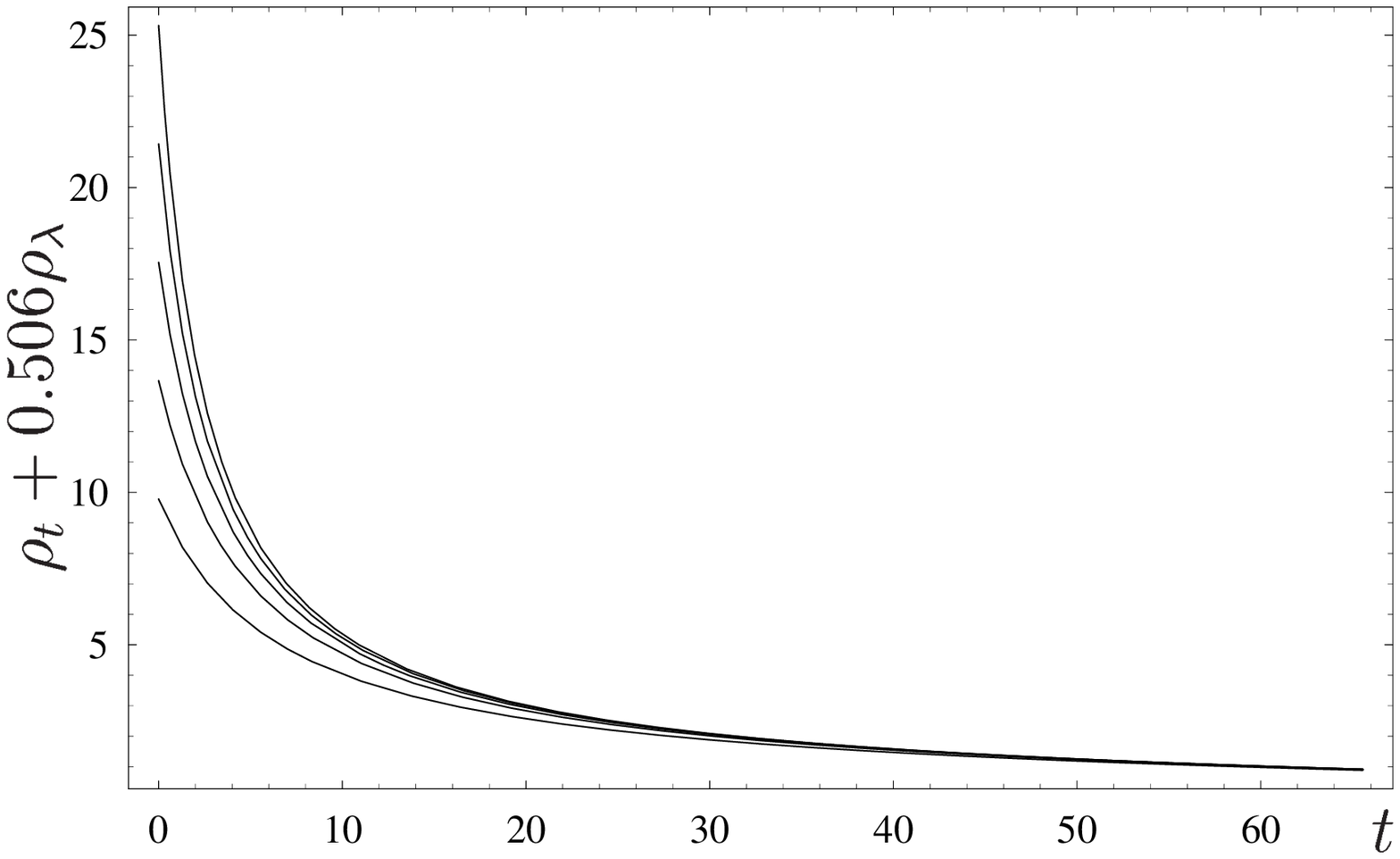}

\vspace{2mm}\hspace*{12mm}{\large\bfseries Fig.3a.}

\vspace{20mm}\includegraphics[height=96mm,keepaspectratio=true]{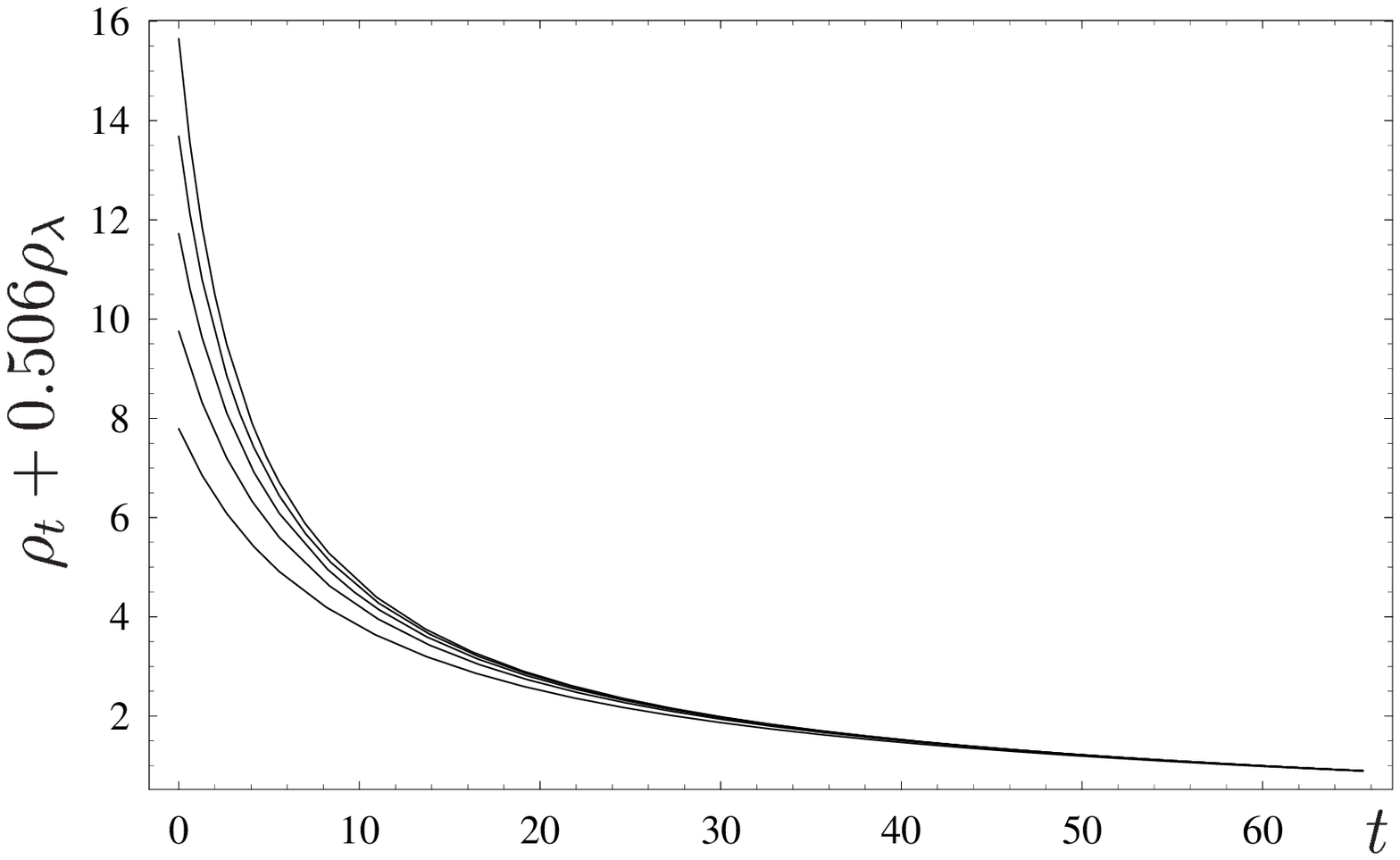}

\vspace{2mm}\hspace*{12mm}{\large\bfseries Fig.3b.}

\end{center}

\newpage

\begin{center}

\includegraphics[height=103mm,keepaspectratio=true]{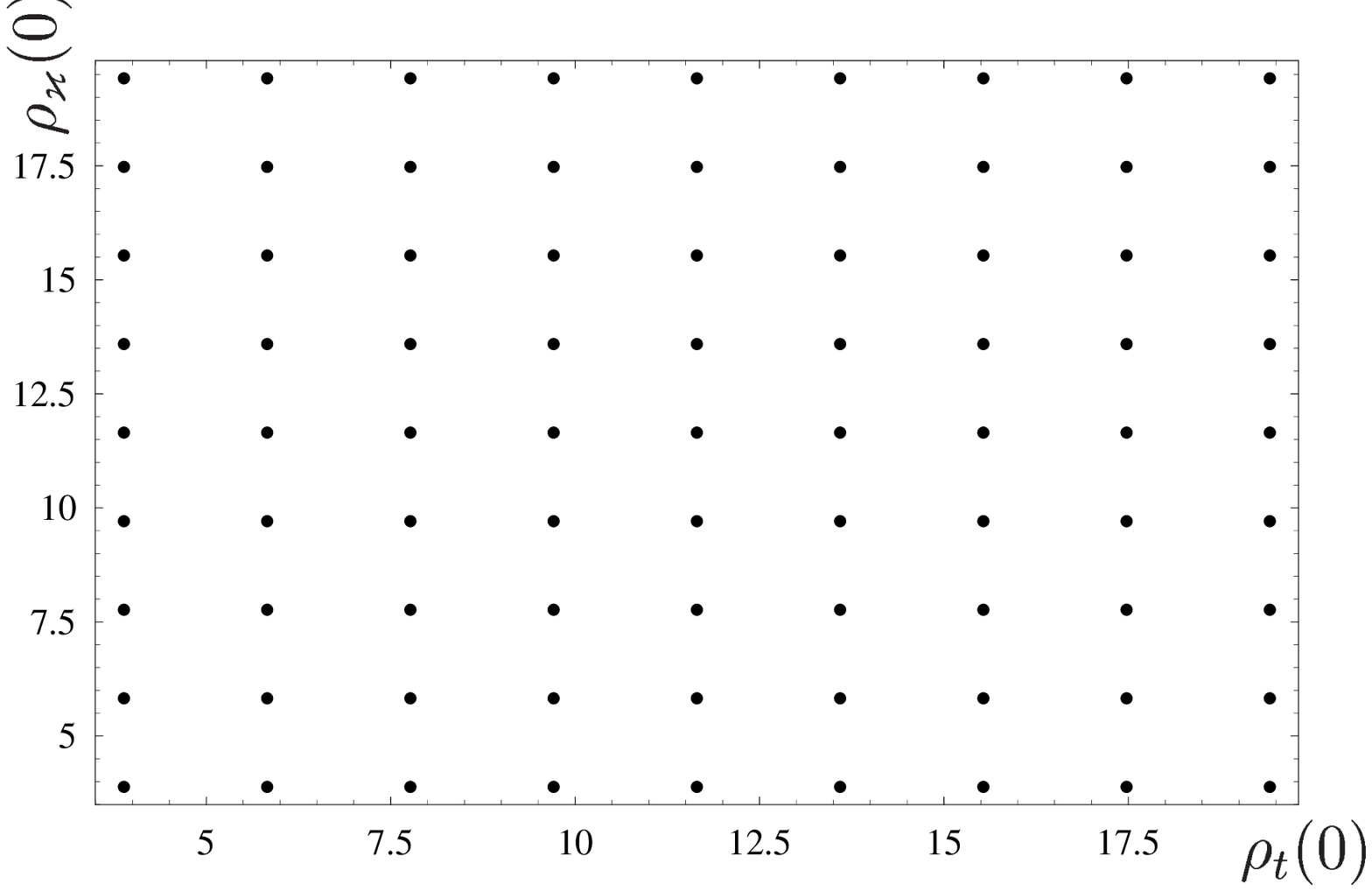}

\vspace{2mm}\hspace*{0mm}{\large\bfseries Fig.4a.}

\vspace{10mm}\includegraphics[height=105mm,keepaspectratio=true]{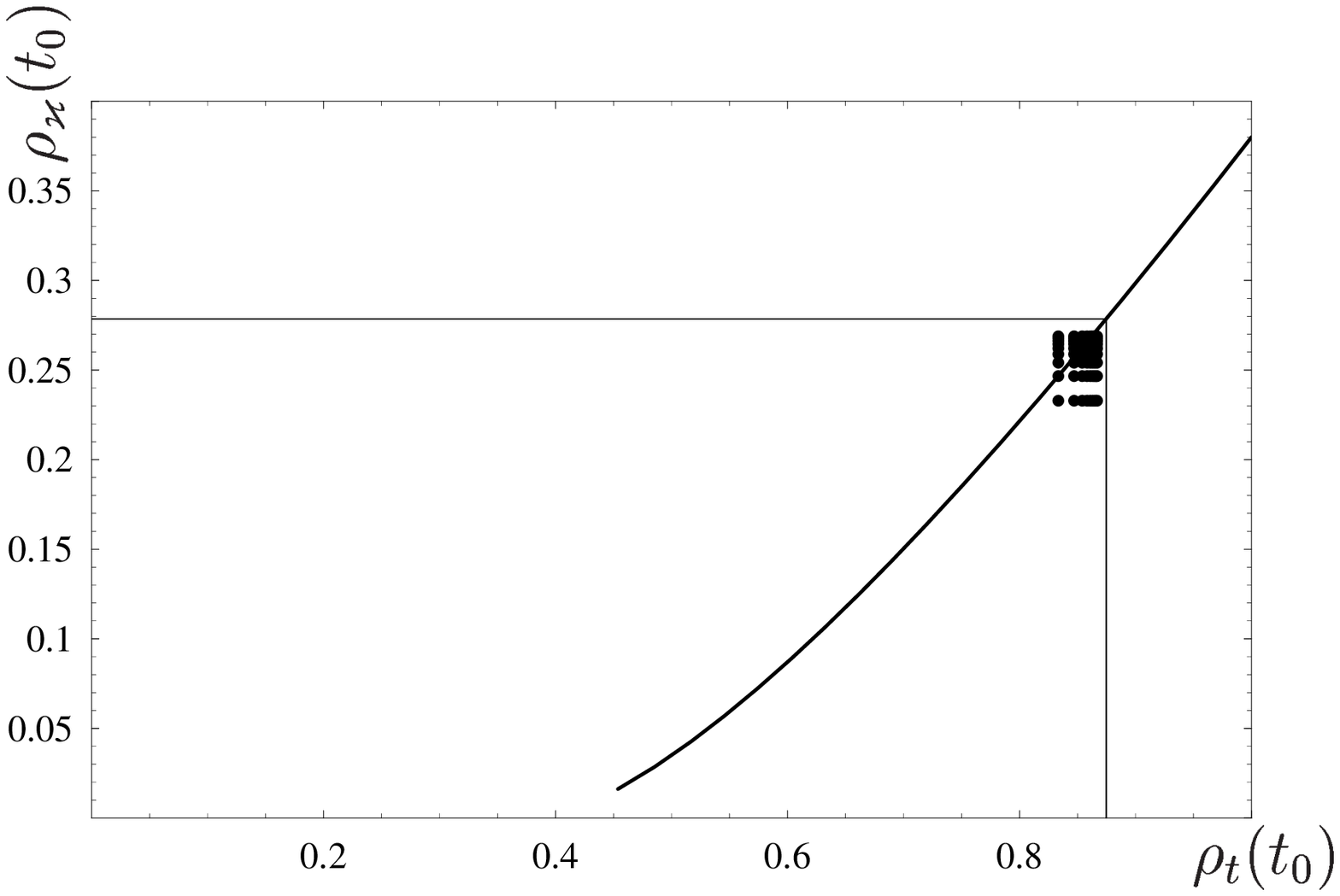}

\vspace{2mm}\hspace*{0mm}{\large\bfseries Fig.4b.}

\end{center}

\newpage

\begin{center}

\includegraphics[height=100mm,keepaspectratio=true]{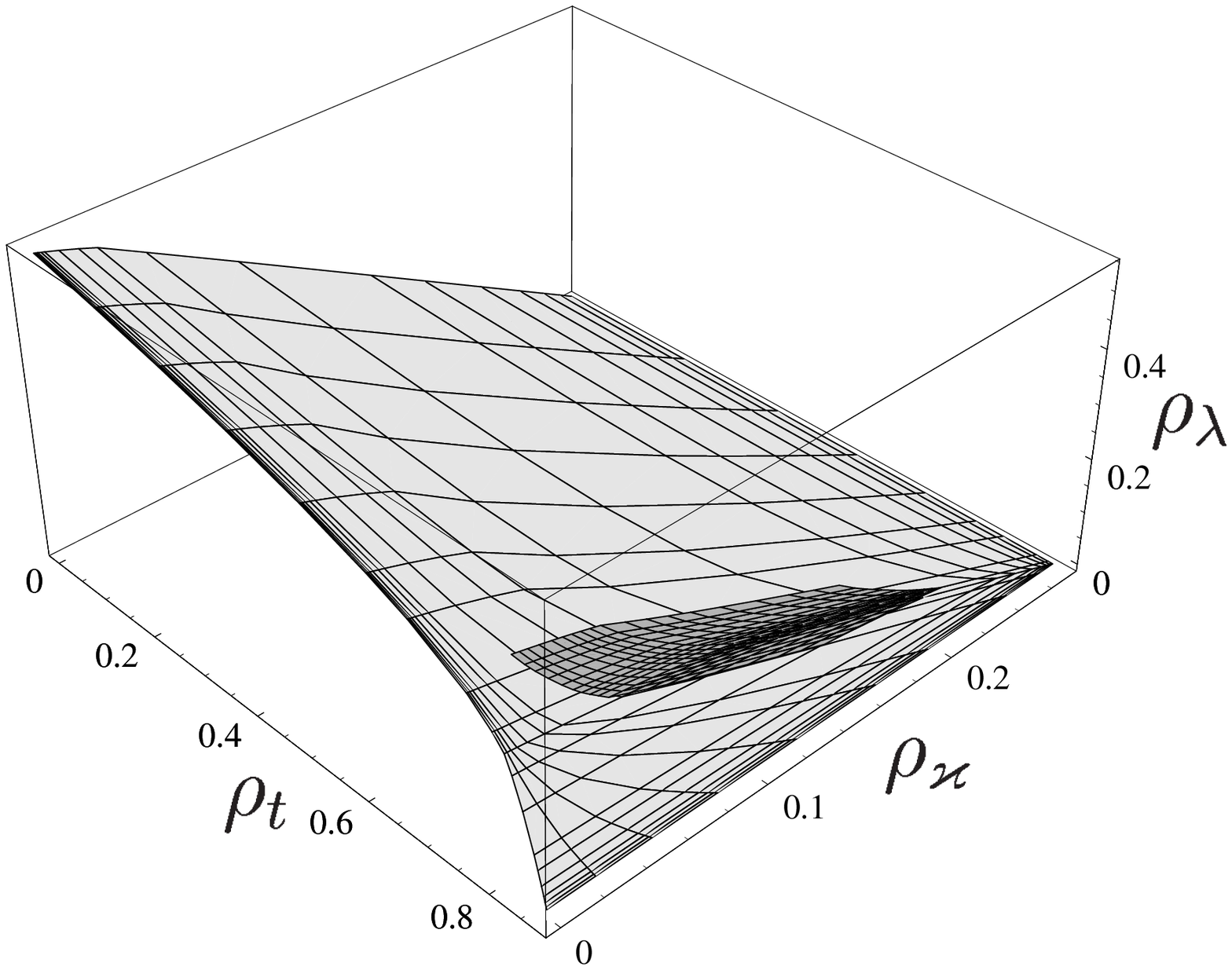}

\vspace{5mm}{\large\bfseries Fig.5a.}

\vspace{15mm}\includegraphics[height=100mm,keepaspectratio=true]{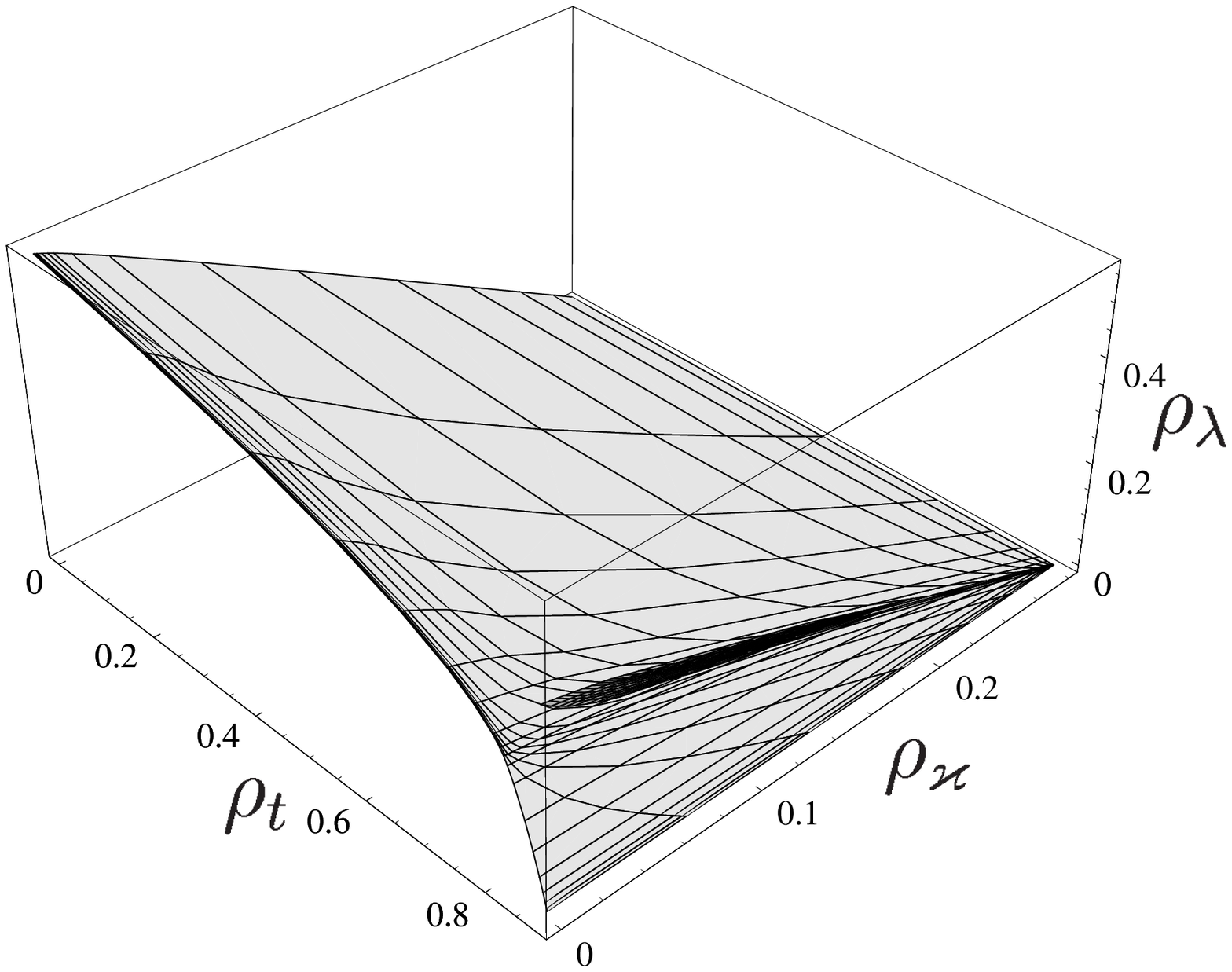}

\vspace{5mm}{\large\bfseries Fig.5b.}

\end{center}

\newpage

\begin{center}

\vspace*{-15mm}
\includegraphics[height=70mm,keepaspectratio=true]{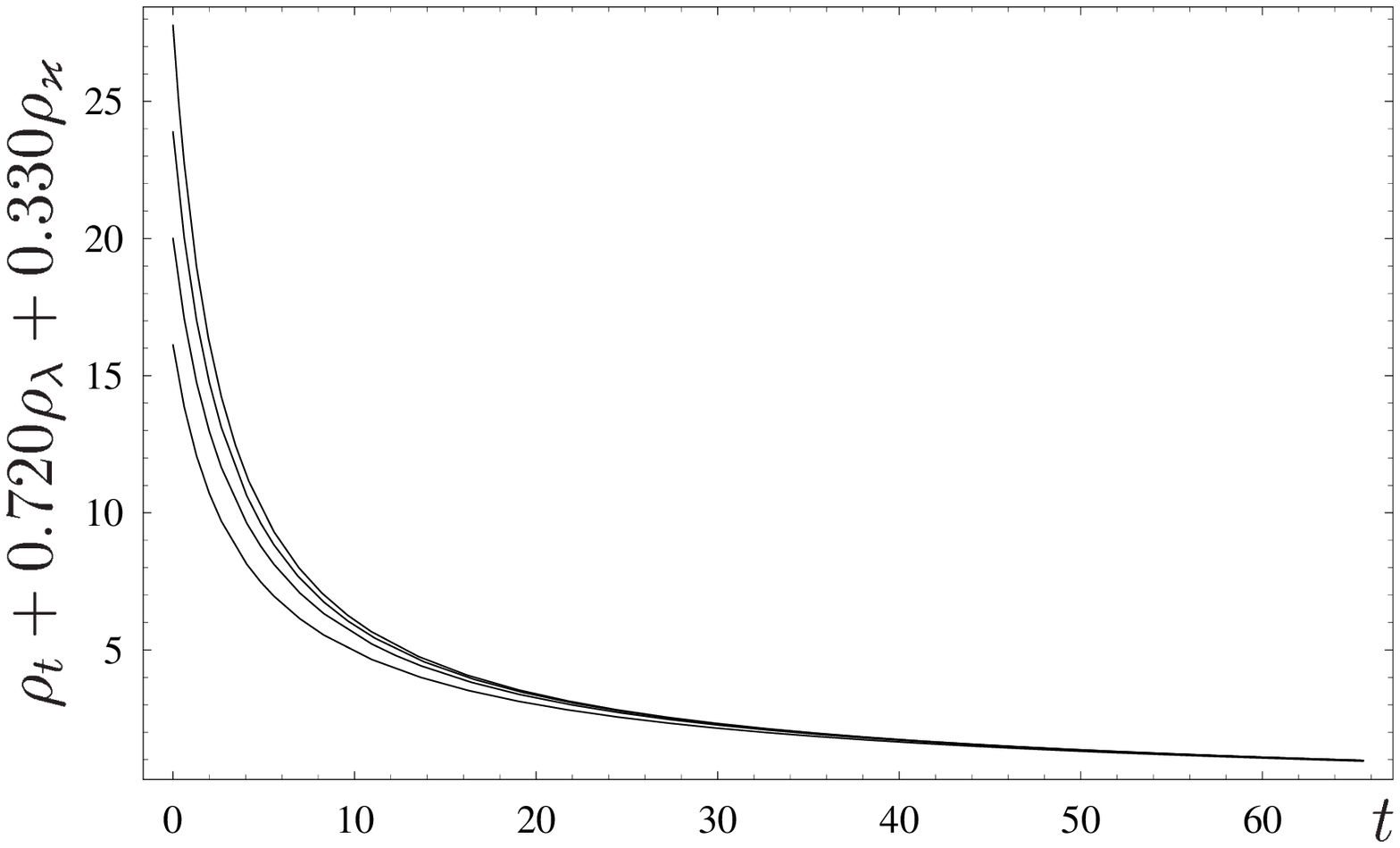}

\vspace{0mm}\hspace*{12mm}{\large\bfseries Fig.6a.}

\vspace{5mm}\includegraphics[height=70mm,keepaspectratio=true]{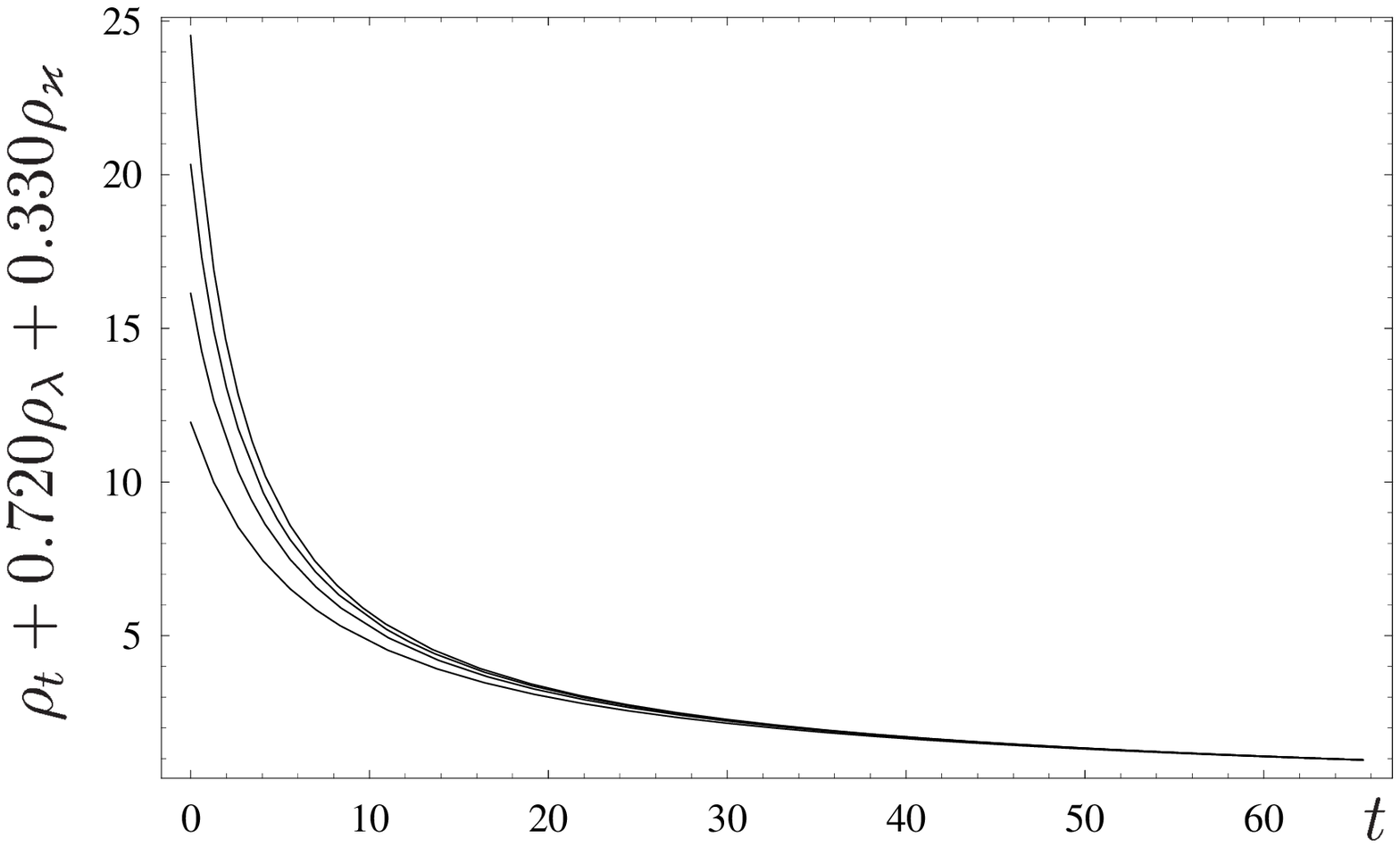}

\vspace{0mm}\hspace*{12mm}{\large\bfseries Fig.6b.}

\vspace{5mm}\includegraphics[height=70mm,keepaspectratio=true]{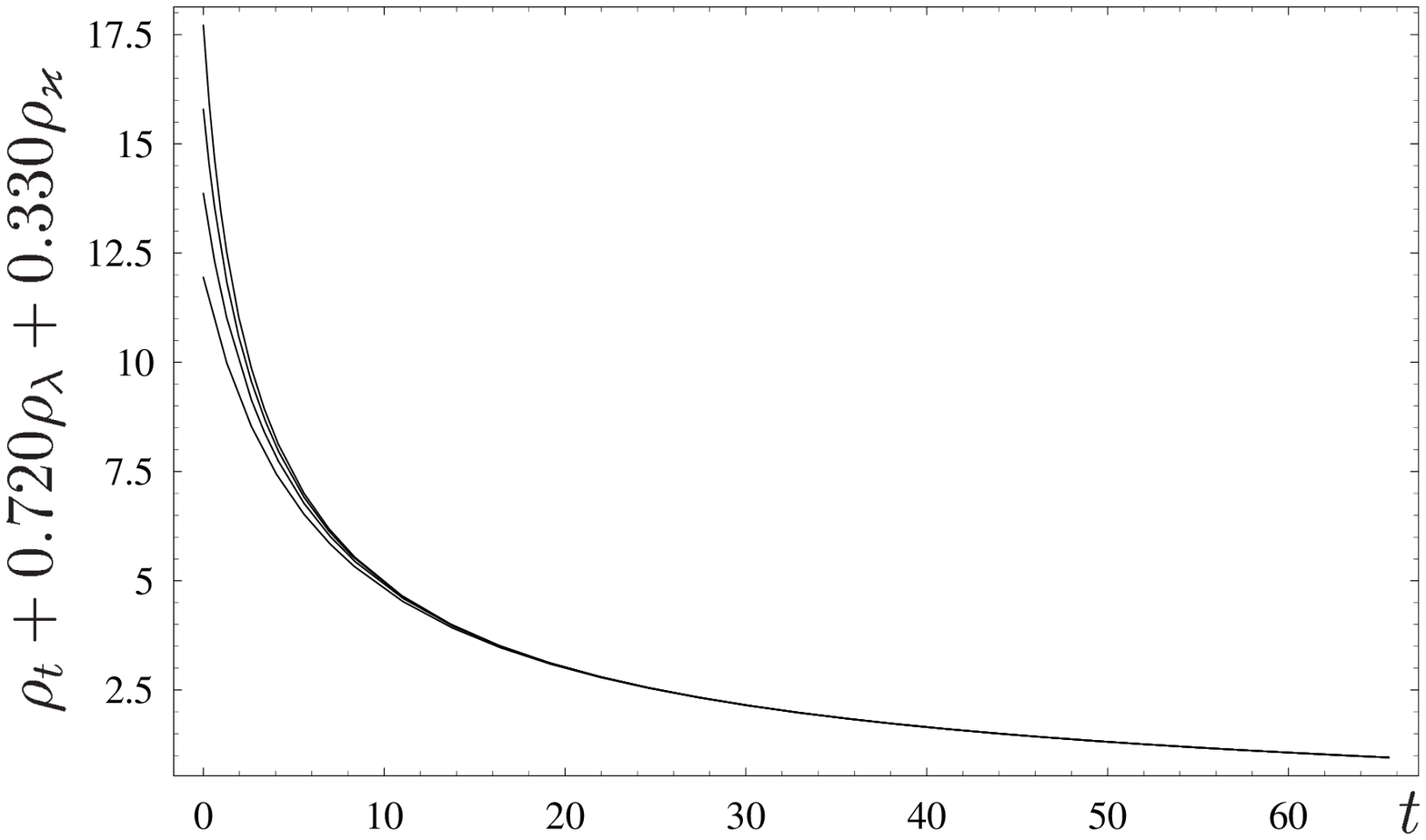}

\vspace{0mm}\hspace*{12mm}{\large\bfseries Fig.6c.}

\end{center}

\begin{landscape}

\begin{center}

\vspace*{-5mm}
\includegraphics[height=144mm,keepaspectratio=true]{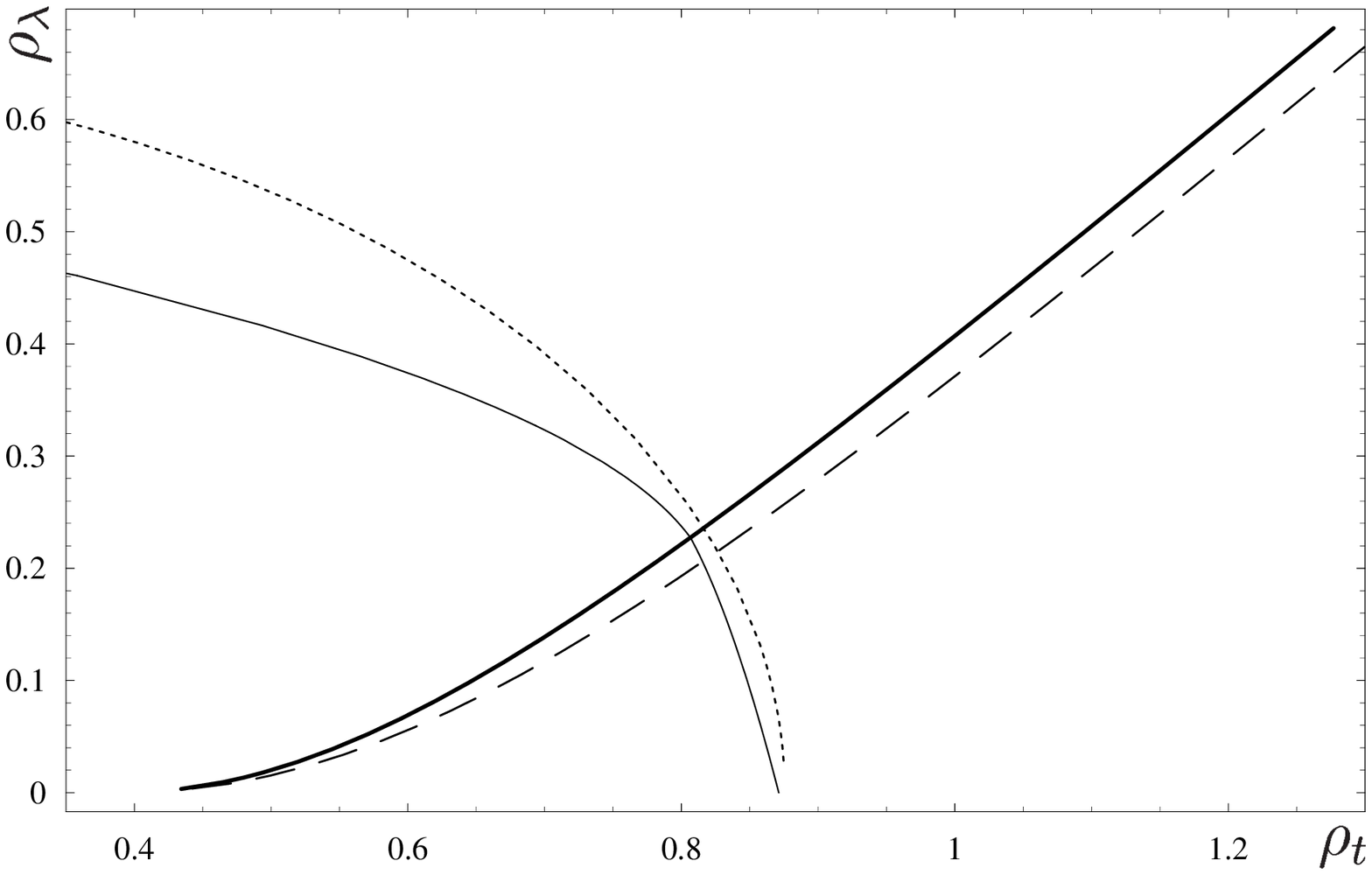}

\vspace{5mm}{\large\bfseries Fig.7.}

\end{center}

\end{landscape}

\begin{landscape}

\begin{center}

\vspace*{-5mm}
\includegraphics[height=144mm,keepaspectratio=true]{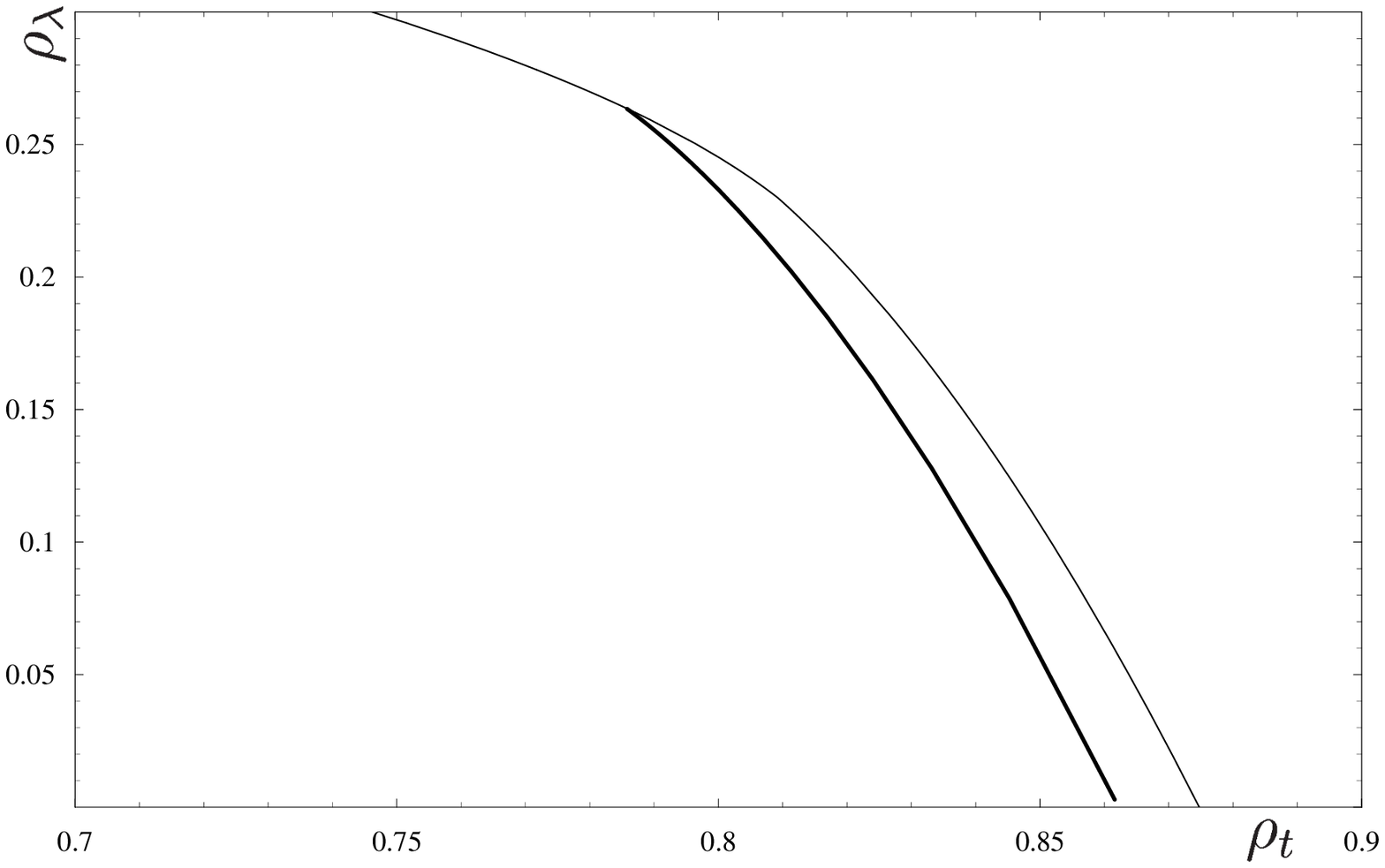}

\vspace{5mm}{\large\bfseries Fig.8.}

\end{center}

\end{landscape}

\end{document}